\documentclass{jkas}

\def\year{2019} 
\def\month{April} 
\def\volume{00} 
\def\issue{0} 
\def\beginpage{1} 
\def\endpage{99} 
\def\received{February 30, 2019} 
\def\accepted{February 31, 2019} 
\date{Received \received; accepted \accepted}

\usepackage[colorlinks,
            ps2pdf,
            breaklinks,
            linkcolor=blue,
            urlcolor=blue,
            anchorcolor=blue,
            menucolor=blue,
            citecolor=blue]{hyperref}

\usepackage{adjustbox}
\usepackage{amsmath}

\title{
Environmental Dependence of Type Ia Supernova Luminosities from the YONSEI Supernova Catalog
}

\author[1, 2]{Young-Lo Kim}
\author[2]{Yijung Kang}
\author[2]{Young-Wook Lee}

\affil[1]{Université de Lyon, F-69622, Lyon, France; Université de Lyon 1, Villeurbanne; CNRS/IN2P3, Institut de Physique Nucléaire de Lyon; \email{y.kim@ipnl.in2p3.fr}}
\affil[2]{Center for Galaxy Evolution Research and Department of Astronomy, Yonsei University, Seoul 03722, Korea}

\def\corrauthor{
Y.-L. Kim
}

\def\runningauthor{
Y.-L. Kim
}

\def\runningtitle{
Environmental Dependence of Type Ia Supernova Luminosities from the YONSEI SN Catalog 
}

\def\keywords{
cosmology: observations --- distance scale --- supernovae: general
}

\def\abstracttext{
There is growing evidence for the dependence of Type Ia supernova (SN Ia) luminosities on their environments. While the impact of this trend on estimating cosmological parameters is widely acknowledged, the origin of this correlation is still under debate. In order to explore this problem, we first construct the YONSEI (YOnsei Nearby Supernova Evolution Investigation) SN catalog. The catalog consists of 1231 spectroscopically confirmed SNe Ia over a wide redshift range ($0.01<z<1.37$) from various SN surveys and includes the light-curve fit data from two independent light-curve fitters of SALT2 and MLCS2k2.  For a sample of 674 host galaxies, we use the stellar mass and the star formation rate data in \citet{Kim2018}. We find that SNe Ia in low-mass and star-forming host galaxies are $0.062\pm0.009$ mag and $0.057\pm0.010$ mag fainter than those in high-mass and passive hosts, after light-curve corrections with SALT2 and MLCS2k2, respectively. When only local environments of SNe Ia (e.g., locally star-forming and locally passive) are considered, this luminosity difference increases to $0.081\pm0.018$ mag for SALT2 and $0.072\pm0.018$ mag for MLCS2k2. Considering the significant difference in the mean stellar population age between the two environments, this result suggests that the origin of environmental dependence is most likely the luminosity evolution of SNe Ia with redshift.
}
\begin{document}
\jkashead 


\section{Introduction\label{sec:intro}}

Observations of distant Type Ia supernovae (SNe Ia) reveal the accelerating expansion of the universe \citep{Riess1998, Perlmutter1999}. The use of SNe Ia as a distance indicator is based on two fundamental ideas. The first idea is that SN Ia luminosities can be empirically standardized \citep{Phillips1993,Tripp1998}. This is obtained by empirical light-curve fitters, such as SALT2 \citep{Guy2007} and MLCS2k2 \citep{Jha2007}, that correct observed ``brighter-slower'' and the ``brighter-bluer'' relations. From this light-curve standardization, the scatter of SN Ia peak luminosity is reduced from $\sim$0.3 mag to $\sim$0.14 mag \citep{Guy2007, Jha2007}.

The other fundamental idea is that the standardization does not evolve with redshift or SN environment. This idea was initially supported by small samples of SNe Ia and their host galaxies: no clear dependence of SNe Ia luminosity, after light-curve shape and color or extinction corrections, on their host morphology was shown \citep{Riess1998, Schmidt1998}. However, more recent studies with larger numbers of SNe Ia have revealed $\sim$2$\sigma$ trends between the SN corrected luminosity and host galaxy morphology, such that SNe Ia in early-type galaxies are brighter than those in late-type galaxies, both at low- and high-redshift ranges \citep{Hicken2009b, Suzuki2012}.

The trends between the corrected luminosity of SNe Ia and host galaxy properties are now intensively investigated, empirically \citep{Gallagher2008, Kelly2010, Lampeitl2010, Sullivan2010, D'Andrea2011, Gupta2011, Childress2013, Johansson2013, Pan2014, Campbell2016, Wolf2016} and theoretically \citep{Hoflich1998, Timmes2003, Kasen2009}. All these studies conclude that  SNe in late-type, star-forming, low-mass, and low-metallicity hosts are $\sim$0.1 mag fainter than those in early-type, passive, high-mass, and high-metallicity galaxies. Even in recent studies using the ``local'' environment at the SN explosion site \citep{Rigault2013, Rigault2015, Rigault2018, Jones2018b, Kim2018, Roman2018}, they also suggest the same result with the similar size of luminosity difference.

Considering the intrinsic scatter of $\sim$0.14 mag on the SN Ia luminosity after ``empirical'' light-curve corrections, $\sim$0.1 mag difference from above host galaxy studies implies that there are ``physical" processes of SNe Ia we have not well understood yet, such as the different population of them and/or their explosion mechanisms. It is therefore important to investigate the origin of the environmental dependence of SN Ia luminosities to understand the underlying physics of an SN, which could lead to an SN as a more accurate standard candle. Some studies suggest that this dependence may arise from differing stellar population properties of their progenitor or host galaxy, such as age and metallicity \citep{Timmes2003, Kasen2009, Johansson2013, Childress2014, Pan2014, Kang2016}. However, most of the previous works are limited by the sample size and the redshift range, because they used SNe only found in a specific survey (e.g., SDSS-II SNe survey or SN Legacy survey) and analyzed with one light-curve fitter (especially with SALT2). Therefore, in order to investigate the origin of luminosity difference and the underlying physics in detail, we require combining survey data to increase the sample size. As a part of the YOnsei Nearby SN Evolution Investigation \citep[YONSEI;][]{Kim2015, Kang2016} project, here we present the YONSEI SN catalog which includes most of the survey data so far to make the sample of 1231 spectroscopically confirmed SNe Ia and 674 host galaxies and has two independent light-curve fitters of SALT2 and MLCS2k2. From this catalog, we show an extended study for the dependence of SN Ia luminosities on the global and local host properties.

\section{Construction of the YONSEI Supernova Catalog\label{sec:catalog}}

For the YONSEI project, we have constructed our own SNe Ia catalog. We employed SALT2 and MLCS2k2 light-curve fitters implemented in the SuperNova ANAlysis software \citep[hereafter SNANA;][]{Kessler2009b} package version 10$\_$34. From this analysis, we have 1231 SNe Ia over the redshift range of $0.01\le z \le 1.37$, which makes this catalog a superset of all SN Ia surveys adopted in the SNANA package.

\subsection{Data Sets\label{sec.2.1}}

SN Ia light-curve data we analyzed are taken from several surveys compiled in the SNANA package except for the Pan-STARRS SN Ia data \citep[hereafter PS, 146 SNe;][]{Rest2014}. \citet{Rest2014} provides the PS light-curves in the SNANA format, so that we add the PS sample to the SNANA light-curve archive. In SNANA, for the  `LOWZ' SNe Ia (see Table~\ref{tab.2.1}), we use the JRK07 compilation of SNe collected from Calan/Tololo \citep[29 SNe;][]{Hamuy1996a, Hamuy1996b}, CfA1 \citep[22 SNe;][]{Riess1999a}, CfA2 \citep[44 SNe;][]{Jha2006}, and other sources \citep[38 SNe;][]{Jha2007}. We also include CfA3 \citep[185 SNe;][]{Hicken2009a}, CfA4 \citep[94 SNe;][]{Hicken2012}, CSP DR1/2 \citep[hereafter CSP, 85 SNe;][]{Contreras2010, Stritzinger2011} for our LOWZ sample. In addition, the full three years SDSS-II SNe suvery \citep[hereafter SDSS, 500 SNe;][]{Sako2018}, ESSENCE survey \citep[60 SNe;][]{Miknaitis2007}, the first three years of Supernova Legacy survey \citep[hereafter SNLS, 281 SNe;][]{Guy2010}, and $HST$ sample \citep[37 SNe;][]{Riess2004, Riess2007} are used for our intermediate- and high-redshift SN samples. In total, 1521 SNe Ia were collected for our light-curve analysis (see Table~\ref{tab.2.2}). If there are same SNe from different surveys, we use the SN which has the most observations \citep[see e.g.,][]{Rest2014}.

\begin{table*}[h]
\centering
\caption{\label{tab.2.1} Contributions to the LOWZ Sample in the YONSEI SN Catalog}
\begin{tabular}{cccccccc}
\toprule
SN & Redshift & Total & \multicolumn{2}{c}{SALT2} & & \multicolumn{2}{c}{MLCS2k2} \\
\cline{4-5} \cline{7-8}
Data & Range & & Cosmology Sample & All Sample & & Cosmology Sample & All Sample \\
\midrule
JRK07  & 0.01-0.09 & 133 & 59  & 69  & & 46  & 69 \\
CFA3   & 0.01-0.07 & 185 & 73  & 89  & & 62  & 96 \\
CFA4   & 0.01-0.08 & 94  & 46  & 59  & & 35  & 66 \\
CSP    & 0.01-0.09 & 85  & 40  & 49  & & 31  & 62 \\
\midrule
LOWZ   & 0.01-0.09 & 497 & 218 & 266 & & 174 & 293 \\
\bottomrule
\end{tabular}
\end{table*}

\begin{table*}[h]
\centering
\caption{\label{tab.2.2} Contributions and Cuts for Each Sample in the YONSEI SN Catalog}
\begin{tabular}{ccccc|ccc}
\toprule
Light-Curve  & SN    & Redshift & Total & Final               & Initial Cut   & Cosmology & Chauvenet's \\
Fitter       & Data  & Range    &       & (=Cosmology sample) & (=All sample) & Cut       & Criterion   \\
\midrule
& \textbf{YONSEI} & 0.01-1.37 & 1521 & \textbf{1049} & 339 & 129 & 4 \\
\cline{2-8}
& LOWZ    & 0.01-0.09 & 497 & 218 & 231 & 45 & 3 \\
& SDSS    & 0.03-0.41 & 500 & 392 & 64  & 43 & 1 \\
SALT2 & PS & 0.03-0.64 & 146 & 108 & 23 & 15 & 0 \\
& ESSENCE & 0.15-0.70 & 60  & 51  & 0   & 9  & 0 \\
& SNLS    & 0.12-1.06 & 281 & 262 & 9   & 10 & 0 \\
& HST     & 0.21-1.37 & 37  & 18  & 12  & 7  & 0\\
\hline
& \textbf{YONSEI} & 0.01-1.37 & 1521 & \textbf{821} & 333 & 362 & 5 \\
\cline{2-8}
& LOWZ    & 0.01-0.09 & 497 & 174 & 204 & 116 & 3 \\
& SDSS    & 0.03-0.41 & 500 & 328 & 82  & 90  & 0 \\
MLCS2k2 & PS & 0.03-0.64 & 146 & 98 & 21 & 27 & 0 \\
& ESSENCE & 0.15-0.70 & 60  & 41  & 1   & 18  & 0 \\
& SNLS    & 0.12-1.06 & 281 & 170 & 9   & 100 & 2 \\
& HST     & 0.21-1.37 & 37  & 10  & 16  & 11  & 0\\
\bottomrule
\end{tabular}
\end{table*}

\subsection{Light-Curve Analysis\label{sec.2.2}}

\subsubsection{SALT2 and MLCS2k2 Light-Curve Fitters Implemented in the SNANA Package\label{sec.2.2.1}}

SNANA package was originally developed for the analysis of SDSS SNe data, and then adopted to simulate and fit the SN Ia light-curves in different surveys and from different telescopes. This package contains a light-curve simulation, a light-curve fitter, and a cosmology fitter for all types of SNe. The primary goal of the SNANA package is to use SNe Ia as distance indicators for the estimation of cosmological parameters. Furthermore, it can also be employed to study the SN rate for the selection efficiency, estimate the non-Ia contamination, and optimize future SN surveys. In the SNANA package, most of the current SN models and non-Ia models are included. For the YONSEI catalog, we employed the most up-to-date SALT2 (version 2.4 from \citet{Betoule2014}) and MLCS2k2 (with $R_{V}$ = 2.2) light-curve fitters. We note that for the MLCS2k2, we select a flat prior allowing negative $A_{V}$ values, and $R_{V}$ = 2.2 for the dust reddening parameter, which is the same value as in \citet{Kessler2009a}.

The purpose of the two light-curve fitters of SALT2 and MLCS2k2 is to determine the luminosity distance of SNe Ia. A basic equation to estimate distances from SNe Ia ($\mu_{SN}$) is parameterized by a linear function of light-curve shape and color or extinction parameters, using the following equation \citep{Tripp1998}:
\begin{equation}
\begin{split}
\mu_{SN} = {}  &  m_{B} - M_{B} + \alpha \times (\textit{light-curve shape}) \\
                        & - \beta \times (\textit{color or extinction}),
\end{split}
\label{eq.DM_SN}
\end{equation}
where $m_{B}$ is an observed rest-frame peak apparent magnitude in $B$-band, $M_{B}$ is an absolute magnitude determined from an SN intrinsic luminosity, and $\alpha$ and $\beta$ are global parameters that characterize the brighter-slower and brighter-bluer relations, respectively. SALT2 and MLCS2k2 fit each observed light-curve for the peak magnitude, the shape parameter ($X_{1}$ for SALT2 and $\Delta$ for MLCS2k2), and the color ($C$ for SALT2) or host galaxy extinction value ($A_{V}$ for MLCS2k2), under an implicit assumption that SNe Ia at low-redshift and those at high-redshift are the same for a given shape and color within the observed scatter. However, these fitters have different ways of estimating model parameters from light-curve data, different approaches to training the models, and different assumptions about the treatment of color variations in SNe Ia. Although SALT2 is the most widely used light-curve fitter in the SN community, MLCS2k2 has a novelty in the context of the investigation of the luminosity evolution of SNe Ia. SALT2 is calibrated by using SNe spanning from the low- to the high-redshift range, while MLCS2k2 is calibrated based solely on SNe at low-redshift \citep[see e.g.,][]{Guy2010, Betoule2014, Jones2015}. At high redshift, young progenitors might be dominant, while young and old progenitors are expected to be mixed at low redshift. Therefore, in SALT2, the possible luminosity evolution effect is more diluted than that in MLCS2k2. In this respect, MLCS2k2 could be more powerful to reveal the luminosity evolution of SNe Ia. In this paper, which also explores the possible luminosity evolution of SNe Ia, we hence use both fitters and analyze them separately, and then compare the results.

\subsubsection{Initial Cut Criteria\label{sec.2.2.2}}

The 1521 SN light-curve data we collected were analyzed by SALT2 and MLCS2k2 fitters with the initial cut criteria. The initial criteria are based on 1) the light-curve data quality, 2) the light-curve fit quality, and 3) the redshift, which are similar to the criteria adopted in \citet{Betoule2014}, \citet{Rest2014}, and \citet{Sako2018}. Of the 1521 SNe, 1182 for SALT2 and 1188 for MLCS2k2 pass this requirement. We detail the selection criteria below and the number of cuts in each data set is listed in Table~\ref{tab.2.2}. 
\newline

\underline{Light-curve data quality criteria:}
\newline

1-1) at least 1 measurement with $-$20 days $<$ t $<$ 

\hspace{0.6cm} $+$10 days, where $t$ is the rest-frame phase rel-

\hspace{0.6cm} ative to the time of maximum light in $B$-band.

1-2) at least 1 measurement with $+$0 days $<$ t $<$ 

\hspace{0.6cm} $+$50 days.

1-3) at least 3 measurements between $–$20 days $<$ 

\hspace{0.6cm} t $<$ $+$50 days.

1-4) 2 or more filters with signal-to-noise ratio $\geq$ 3.
\newline

\underline{Light-curve fit quality and the redshift criteria:}
\newline

2-1) $P_{fit}$ $\geq$ 0.01, where $P_{fit}$ is the SNANA light-

\hspace{0.6cm} curve fit probability based on the $\chi^{2}$ per degree

\hspace{0.6cm} of freedom.

2-2) Visual inspection for the SN light-curve fit.

3-1) $z \geq 0.01$, in order to include SNe Ia in the

\hspace{0.6cm}Hubble flow.
\newline

\citet{Conley2011} pointed out that the cut using an automated quality of fit based on $\chi^{2}$ statistics (e.g., $P_{fit}$) is often misleading, especially for the LOWZ sample. Many SNe in the LOWZ sample have the occasional outlying measurements which have little or no effect on the light-curve fitting, but derive a large value of $\chi^{2}$ of the fit. In order to prevent this situation, we therefore have performed a visual inspection, together with the value of $P_{fit}$, to check the light-curve fit quality.

\subsubsection{Bias Correction and Error Analysis\label{sec.2.2.3}}

\begin{figure}
\centering
\includegraphics[angle=-90,width=\columnwidth]{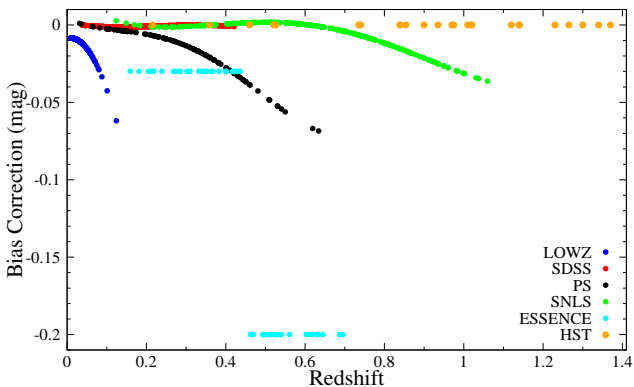}
\caption[Bias correction for each SN in the YONSEI SN Catalog as a function of redshift]{Bias correction for each SN in the YONSEI SN Catalog as a function of redshift. We subtracted this value from all of the rest-frame peak magnitudes in $B$-band for SALT2 and all of the distance modulus for MLCS2k2 in our sample.}
\label{fig.2.3}
\end{figure}

Before we determine the best-fit cosmological parameters from our sample, we should first consider the selection bias and the uncertainties in distance modulus. SN samples from flux-limited surveys are affected by the Malmquist bias \citep{Malmquist1936} or the selection bias: intrinsically brighter and more slowly declining SNe Ia would less suffer from the selection effect and stay longer above the detection threshold than intrinsically fainter and fast declining SNe, and hence be easier to observe. This leads to creating a false result of increasing luminosity of SNe Ia at a given distance. Consequently, the bias has an effect on the estimating distance, such that the determined SN luminosity distance might be closer than the true value. Therefore, we should correct the Malmquist bias to obtain accurate cosmological parameters. Since the effect of the bias differs from survey to survey, we need to explicitly correct it for our analysis according to each survey. We take correction terms for LOWZ, SDSS, and SNLS samples from those calculated by \citet{Betoule2014}, which has a very similar sample to our catalog. For the PS sample, we take the term from \citet{Rest2014}, and from \citet{Wood-Vasey2007} for the ESSENCE sample. As \citet{Strolger2004} argued that the HST searches are sufficiently deep to suffer from little or no Malmquist bias out to the maximum redshift where SN discovered, we do not perform the bias correction for the HST sample (see also \citealp{Conley2011}). We then interpolate the correction value for each SN at a given redshift (Figure~\ref{fig.2.3}), and this value is subtracted from all of the rest-frame peak magnitudes in the $B$-band for SALT2 and all of the distance moduli for MLCS2k2 in our catalog.

The total uncertainty is generally computed by the propagation of statistical and systematic uncertainties. The systematic uncertainties are associated with the calibration, the light-curve model uncertainty, the host dependency, and so on. In this paper, our purpose is to investigate one of the systematic uncertainties, the host dependency, so here we consider only the statistical uncertainties.

Following \citet{Conley2011} and \citet{Betoule2014}, the (statistical) uncertainty for each SN is propagated in the polynomial form given by

\begin{equation}
\sigma_{stat}^{2} = \sigma_{fit}^{2} + \sigma_{z}^{2} + \sigma_{lens}^{2} + \sigma_{int}^{2},
\label{eq.sigma.stat}
\end{equation}

\hspace{-0.8cm} where $\sigma_{fit}$ is the error on the fitted light-curve parameters estimated from SALT2 and MLCS2k2, $\sigma_{z}$ accounts for the redshift determination uncertainty, $\sigma_{lens}$ represents the statistical variation of magnitudes caused by the gravitational lensing, and $\sigma_{int}$ is the intrinsic scatter of SNe Ia to make a reduced $\chi^{2}$ ($\chi_{red}^{2}$; the $\chi^2$ per degree of freedom) a unity when we determine the best-fit cosmology. Our approximation of the redshift uncertainty follows Equation (5) of \citet{Conley2011}, and the random, uncorrelated scatter due to the lensing follows the suggestion of \citet{Jonsson2010}: $\sigma_{lens} = 0.055 \times z$. The intrinsic scatter is not included in our analysis, as we are searching this extra scatter through the systematic variation in SN Ia luminosity on the host galaxy properties \citep{D'Andrea2011, Gupta2011, Pan2014}.

\subsection{YONSEI Supernova Catalog and Systematic Tests\label{sec.2.3}}

\begin{figure}
\centering
\includegraphics[angle=-90,width=\columnwidth]{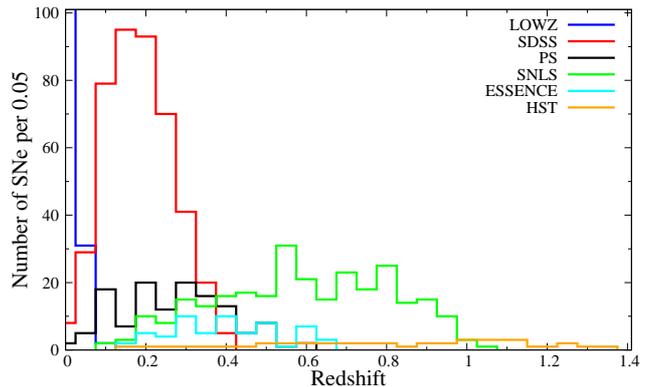}
\caption[Redshift distribution of the YONSEI All sample]{Redshift distribution of the YONSEI All sample. Histograms are colored by surveys. The first bin of the LOWZ sample contains 268 SNe Ia.}
\label{fig.2.4}
\end{figure}

After applying the bias correction and the error propagation for the uncertainty, now we have the YONSEI SN Catalog (see Table~\ref{tab.A.Catalog}). This catalog provides a rest-frame peak magnitude in $B$-band or distance modulus, a light-curve shape parameter, and a color or host extinction value for each SN. Because both normal and peculiar SNe Ia are included, we call this catalog as the YONSEI `All' sample to distinguish from the YONSEI `Cosmology' sample which only has normal SNe described in next Section. The redshift distribution of the YONSEI All sample is shown in Figure~\ref{fig.2.4}. 

\begin{table*}
\begin{center}
\caption{\label{tab.A.Catalog} YONSEI Supernova Catalog with Host Properties}
\begin{adjustbox}{width=1.1\textwidth,center=\textwidth}
\begin{tabular}{lccccccccccccccccccccccc}
\toprule
     &        &           & \multicolumn{6}{c}{SALT2}                       && \multicolumn{6}{c}{MLCS2k2}                        &            & \multicolumn{3}{c}{Host Mass}              && \multicolumn{3}{c}{Global sSFR} \\
\cline{4-9} \cline{11-16} \cline{18-20} \cline{22-24}
Name & Survey & z$_{CMB}$ & $m_{B}$ & Error & $X_{1}$ & Error & $C$ & Error && $\mu_{SN}$ & Error & $\Delta$ & Error & $A_{V}$ & Error & Morphology & log($M_{stellar}$) & $-\delta$ & $+\delta$ && log($sSFR$) & $-\delta$ & $+\delta$ \\
     &        &           & (mag)   &       &         &       &     &       && (mag)      &       &          &       & (mag)   &       &            & (M$_{\odot}$)      &           &           && (yr$^{-1}$) &           &           \\

\midrule
1990O	&	LOWZ	&	0.0306 	&	16.230 	&	0.035 	&	0.541 	&	0.180 	&	-0.056 	&	0.030 	&	&	35.839 	&	0.121 	&	-0.202 	&	0.074 	&	0.067 	&	0.090 	&	Sa	&	...	&	...	&	...	&	&	...	&	...	&	...	\\
1990af	&	LOWZ	&	0.0502 	&	17.777 	&	0.032 	&	-2.132 	&	0.141 	&	-0.043 	&	0.044 	&	&	36.846 	&	0.111 	&	0.651 	&	0.156 	&	-0.195 	&	0.161 	&	S0	&	...	&	...	&	...	&	&	...	&	...	&	...	\\
1991ag	&	LOWZ	&	0.0139 	&	14.438 	&	0.046 	&	0.843 	&	0.144 	&	-0.050 	&	0.030 	&	&	34.082 	&	0.111 	&	-0.228 	&	0.068 	&	0.050 	&	0.104 	&	Sd	&	9.07 	&	0.03 	&	0.03 	&	&	-8.66 	&	0.06 	&	0.08 	\\
1992P	&	LOWZ	&	0.0263 	&	16.074 	&	0.034 	&	0.217 	&	0.277 	&	-0.073 	&	0.041 	&	&	35.596 	&	0.118 	&	-0.171 	&	0.086 	&	0.120 	&	0.078 	&	Sb	&	10.34 	&	0.10 	&	0.14 	&	&	-9.98 	&	0.88 	&	0.63 	\\
1992ae	&	LOWZ	&	0.0748 	&	18.459 	&	0.043 	&	-0.729 	&	0.153 	&	-0.024 	&	0.051 	&	&	37.703 	&	0.168 	&	0.128 	&	0.114 	&	0.116 	&	0.147 	&	E	&	...	&	...	&	...	&	&	...	&	...	&	...	\\
1992ag	&	LOWZ	&	0.0259 	&	16.372 	&	0.036 	&	-0.640 	&	0.094 	&	0.212 	&	0.041 	&	&	35.095 	&	0.107 	&	0.084 	&	0.073 	&	0.465 	&	0.062 	&	Sc	&	10.02 	&	0.10 	&	1.04 	&	&	-9.88 	&	1.23 	&	0.93 	\\
1992al	&	LOWZ	&	0.0141 	&	14.469 	&	0.033 	&	-0.256 	&	0.083 	&	-0.112 	&	0.026 	&	&	34.109 	&	0.074 	&	-0.025 	&	0.064 	&	-0.062 	&	0.053 	&	Sc	&	...	&	...	&	...	&	&	...	&	...	&	...	\\
1992aq	&	LOWZ	&	0.1009 	&	19.317 	&	0.041 	&	-1.375 	&	0.219 	&	-0.089 	&	0.056 	&	&	...	&	...	&	...	&	...	&	...	&	...	&	Sa	&	...	&	...	&	...	&	&	...	&	...	&	...	\\
1992bc	&	LOWZ	&	0.0198 	&	15.121 	&	0.031 	&	0.881 	&	0.072 	&	-0.115 	&	0.026 	&	&	34.960 	&	0.050 	&	-0.261 	&	0.043 	&	-0.078 	&	0.040 	&	Sc	&	9.72 	&	0.53 	&	0.70 	&	&	-9.62 	&	0.94 	&	1.12 	\\
1992bh	&	LOWZ	&	0.0451 	&	17.625 	&	0.035 	&	0.003 	&	0.148 	&	0.044 	&	0.044 	&	&	36.840 	&	0.076 	&	-0.102 	&	0.068 	&	0.276 	&	0.060 	&	Sbc	&	...	&	...	&	...	&	&	...	&	...	&	...	\\
1992bk	&	LOWZ	&	0.0579 	&	18.119 	&	0.045 	&	-1.668 	&	0.150 	&	-0.091 	&	0.052 	&	&	37.146 	&	0.159 	&	0.644 	&	0.248 	&	-0.208 	&	0.163 	&	E	&	...	&	...	&	...	&	&	...	&	...	&	...	\\
1992bl	&	LOWZ	&	0.0429 	&	17.333 	&	0.040 	&	-1.703 	&	0.085 	&	-0.050 	&	0.044 	&	&	36.533 	&	0.122 	&	0.493 	&	0.138 	&	-0.206 	&	0.109 	&	Sa	&	11.81 	&	0.65 	&	0.46 	&	&	-10.85 	&	0.71 	&	1.58 	\\
1992bo	&	LOWZ	&	0.0181 	&	15.782 	&	0.033 	&	-1.980 	&	0.064 	&	-0.045 	&	0.028 	&	&	...	&	...	&	...	&	...	&	...	&	...	&	S0	&	12.13 	&	1.20 	&	0.14 	&	&	-11.18 	&	0.82 	&	1.69 	\\
1992bp	&	LOWZ	&	0.0789 	&	18.310 	&	0.031 	&	-0.952 	&	0.133 	&	-0.104 	&	0.048 	&	&	37.846 	&	0.087 	&	0.089 	&	0.105 	&	-0.061 	&	0.074 	&	E/S0	&	...	&	...	&	...	&	&	...	&	...	&	...	\\
1992br	&	LOWZ	&	0.0878 	&	19.202 	&	0.076 	&	-2.389 	&	0.198 	&	-0.141 	&	0.071 	&	&	38.418 	&	0.198 	&	0.552 	&	0.140 	&	-0.262 	&	0.224 	&	E	&	...	&	...	&	...	&	&	...	&	...	&	...	\\
1992bs	&	LOWZ	&	0.0634 	&	18.299 	&	0.040 	&	-0.241 	&	0.129 	&	-0.027 	&	0.046 	&	&	37.544 	&	0.134 	&	-0.051 	&	0.082 	&	0.208 	&	0.118 	&	Sc	&	...	&	...	&	...	&	&	...	&	...	&	...	\\
1993B	&	LOWZ	&	0.0707 	&	18.467 	&	0.048 	&	-0.331 	&	0.166 	&	0.047 	&	0.051 	&	&	37.668 	&	0.113 	&	-0.022 	&	0.093 	&	0.172 	&	0.091 	&	Sb	&	...	&	...	&	...	&	&	...	&	...	&	...	\\
1993H	&	LOWZ	&	0.0248 	&	16.656 	&	0.035 	&	-2.042 	&	0.060 	&	0.075 	&	0.029 	&	&	...	&	...	&	...	&	...	&	...	&	...	&	Sab	&	10.51 	&	0.38 	&	0.56 	&	&	-8.63 	&	1.74 	&	0.49 	\\
1993O	&	LOWZ	&	0.0519 	&	17.640 	&	0.032 	&	-0.614 	&	0.100 	&	-0.082 	&	0.043 	&	&	37.114 	&	0.073 	&	0.038 	&	0.063 	&	0.031 	&	0.054 	&	E/S0	&	...	&	...	&	...	&	&	...	&	...	&	...	\\
1993ag	&	LOWZ	&	0.0500 	&	17.834 	&	0.034 	&	-0.832 	&	0.136 	&	0.057 	&	0.045 	&	&	37.001 	&	0.098 	&	0.098 	&	0.086 	&	0.176 	&	0.068 	&	E/S0	&	...	&	...	&	...	&	&	...	&	...	&	...	\\
1994M	&	LOWZ	&	0.0243 	&	16.294 	&	0.035 	&	-1.405 	&	0.090 	&	0.055 	&	0.028 	&	&	35.235 	&	0.068 	&	0.293 	&	0.079 	&	0.141 	&	0.073 	&	E	&	11.04 	&	0.11 	&	0.19 	&	&	-10.88 	&	1.12 	&	1.01 	\\
1994S	&	LOWZ	&	0.0160 	&	14.778 	&	0.035 	&	0.376 	&	0.152 	&	-0.079 	&	0.027 	&	&	34.429 	&	0.076 	&	-0.124 	&	0.081 	&	-0.024 	&	0.068 	&	Sab	&	10.50 	&	0.11 	&	0.03 	&	&	-9.68 	&	0.04 	&	0.14 	\\
1994T	&	LOWZ	&	0.0357 	&	17.254 	&	0.032 	&	-1.439 	&	0.112 	&	0.055 	&	0.030 	&	&	...	&	...	&	...	&	...	&	...	&	...	&	S0/a	&	...	&	...	&	...	&	&	...	&	...	&	...	\\
1995ac	&	LOWZ	&	0.0488 	&	17.078 	&	0.029 	&	0.788 	&	0.089 	&	-0.045 	&	0.026 	&	&	36.494 	&	0.053 	&	-0.303 	&	0.047 	&	0.276 	&	0.038 	&	Sa	&	...	&	...	&	...	&	&	...	&	...	&	...	\\
1995ak	&	LOWZ	&	0.0220 	&	16.024 	&	0.039 	&	-1.037 	&	0.109 	&	0.075 	&	0.031 	&	&	34.912 	&	0.086 	&	0.038 	&	0.067 	&	0.372 	&	0.077 	&	Sbc	&	...	&	...	&	...	&	&	...	&	...	&	...	\\
1996C	&	LOWZ	&	0.0275 	&	16.643 	&	0.034 	&	0.741 	&	0.132 	&	0.061 	&	0.029 	&	&	35.867 	&	0.070 	&	-0.162 	&	0.051 	&	0.265 	&	0.062 	&	Sa	&	9.99 	&	0.03 	&	0.17 	&	&	-9.98 	&	0.39 	&	0.60 	\\
1996ab	&	LOWZ	&	0.1242 	&	19.557 	&	0.045 	&	-0.018 	&	0.375 	&	-0.156 	&	0.055 	&	&	...	&	...	&	...	&	...	&	...	&	...	&	S	&	...	&	...	&	...	&	&	...	&	...	&	...	\\
1996bl	&	LOWZ	&	0.0348 	&	16.710 	&	0.031 	&	-0.110 	&	0.110 	&	-0.001 	&	0.027 	&	&	36.048 	&	0.079 	&	-0.139 	&	0.065 	&	0.237 	&	0.061 	&	Sc	&	...	&	...	&	...	&	&	...	&	...	&	...	\\
1996bv	&	LOWZ	&	0.0167 	&	15.341 	&	0.032 	&	0.746 	&	0.103 	&	0.158 	&	0.026 	&	&	...	&	...	&	...	&	...	&	...	&	...	&	Scd	&	10.15 	&	0.98 	&	0.30 	&	&	-9.76 	&	0.89 	&	1.33 	\\
1997E	&	LOWZ	&	0.0133 	&	15.114 	&	0.029 	&	-1.702 	&	0.087 	&	0.016 	&	0.022 	&	&	34.152 	&	0.063 	&	0.244 	&	0.103 	&	0.182 	&	0.073 	&	S0	&	11.52 	&	0.09 	&	0.81 	&	&	-10.03 	&	0.98 	&	1.63 	\\
1997Y	&	LOWZ	&	0.0166 	&	15.329 	&	0.029 	&	-0.940 	&	0.138 	&	-0.015 	&	0.023 	&	&	34.583 	&	0.081 	&	0.059 	&	0.077 	&	0.137 	&	0.059 	&	Sb	&	10.42 	&	0.28 	&	0.01 	&	&	-10.26 	&	0.67 	&	1.16 	\\
1997dg	&	LOWZ	&	0.0297 	&	16.837 	&	0.027 	&	-0.348 	&	0.173 	&	-0.038 	&	0.023 	&	&	36.183 	&	0.076 	&	-0.048 	&	0.095 	&	0.149 	&	0.064 	&	S	&	...	&	...	&	...	&	&	...	&	...	&	...	\\
1997do	&	LOWZ	&	0.0105 	&	14.383 	&	0.034 	&	0.384 	&	0.111 	&	0.092 	&	0.022 	&	&	33.537 	&	0.101 	&	-0.135 	&	0.093 	&	0.338 	&	0.071 	&	Sbc	&	9.30 	&	0.14 	&	0.09 	&	&	-8.65 	&	0.24 	&	0.24 	\\
1998ab	&	LOWZ	&	0.0279 	&	16.100 	&	0.027 	&	0.220 	&	0.073 	&	0.054 	&	0.022 	&	&	35.204 	&	0.067 	&	-0.162 	&	0.054 	&	0.403 	&	0.045 	&	Sbc	&	10.59 	&	0.04 	&	0.23 	&	&	-9.67 	&	0.64 	&	0.72 	\\
1998bp	&	LOWZ	&	0.0102 	&	15.308 	&	0.034 	&	-2.492 	&	0.092 	&	0.194 	&	0.024 	&	&	...	&	...	&	...	&	...	&	...	&	...	&	E	&	...	&	...	&	...	&	&	...	&	...	&	...	\\
1998dx	&	LOWZ	&	0.0537 	&	17.542 	&	0.036 	&	-1.575 	&	0.288 	&	-0.129 	&	0.027 	&	&	37.018 	&	0.074 	&	0.371 	&	0.154 	&	-0.230 	&	0.154 	&	Sb	&	11.72 	&	0.55 	&	0.87 	&	&	-12.00 	&	0.00 	&	4.00 	\\
1998ef	&	LOWZ	&	0.0167 	&	14.846 	&	0.030 	&	-1.122 	&	0.106 	&	-0.070 	&	0.023 	&	&	34.098 	&	0.126 	&	0.231 	&	0.169 	&	0.029 	&	0.108 	&	S	&	...	&	...	&	...	&	&	...	&	...	&	...	\\
1998eg	&	LOWZ	&	0.0235 	&	16.114 	&	0.027 	&	-0.507 	&	0.210 	&	-0.005 	&	0.023 	&	&	35.335 	&	0.078 	&	0.021 	&	0.126 	&	0.209 	&	0.086 	&	Scd	&	11.32 	&	0.75 	&	0.47 	&	&	-11.10 	&	0.69 	&	2.94 	\\
1999aa	&	LOWZ	&	0.0153 	&	14.720 	&	0.027 	&	1.190 	&	0.034 	&	-0.091 	&	0.020 	&	&	34.474 	&	0.034 	&	-0.304 	&	0.022 	&	0.015 	&	0.024 	&	Sc	&	10.72 	&	0.10 	&	0.24 	&	&	-10.23 	&	0.32 	&	1.39 	\\
1999aw	&	LOWZ	&	0.0392 	&	16.797 	&	0.029 	&	2.264 	&	0.075 	&	-0.038 	&	0.027 	&	&	...	&	...	&	...	&	...	&	...	&	...	&	...	&	...	&	...	&	...	&	&	...	&	...	&	...	\\
1999cc	&	LOWZ	&	0.0315 	&	16.779 	&	0.025 	&	-1.546 	&	0.073 	&	-0.012 	&	0.022 	&	&	35.892 	&	0.050 	&	0.268 	&	0.083 	&	0.108 	&	0.058 	&	Sc	&	10.99 	&	0.05 	&	0.04 	&	&	-10.02 	&	0.27 	&	0.42 	\\
1999cp	&	LOWZ	&	0.0104 	&	13.947 	&	0.035 	&	0.334 	&	0.083 	&	-0.087 	&	0.026 	&	&	33.556 	&	0.099 	&	-0.116 	&	0.119 	&	0.014 	&	0.086 	&	Scd	&	9.48 	&	0.09 	&	0.29 	&	&	-8.65 	&	0.39 	&	0.40 	\\
1999dk	&	LOWZ	&	0.0139 	&	14.867 	&	0.030 	&	0.555 	&	0.109 	&	0.048 	&	0.022 	&	&	34.174 	&	0.067 	&	-0.258 	&	0.049 	&	0.283 	&	0.058 	&	Sc	&	10.20 	&	0.16 	&	0.09 	&	&	-9.77 	&	0.37 	&	0.52 	\\
1999dq	&	LOWZ	&	0.0135 	&	14.375 	&	0.028 	&	0.889 	&	0.034 	&	0.021 	&	0.020 	&	&	33.687 	&	0.033 	&	-0.315 	&	0.023 	&	0.369 	&	0.022 	&	Sc	&	10.78 	&	0.06 	&	0.21 	&	&	-9.77 	&	0.83 	&	0.61 	\\
1999ef	&	LOWZ	&	0.0380 	&	17.075 	&	0.056 	&	0.297 	&	0.161 	&	-0.042 	&	0.027 	&	&	36.621 	&	0.105 	&	-0.119 	&	0.125 	&	-0.075 	&	0.128 	&	Scd	&	...	&	...	&	...	&	&	...	&	...	&	...	\\
1999ej	&	LOWZ	&	0.0128 	&	15.339 	&	0.052 	&	-1.485 	&	0.223 	&	-0.032 	&	0.031 	&	&	34.506 	&	0.104 	&	0.321 	&	0.181 	&	-0.045 	&	0.144 	&	S0/a	&	...	&	...	&	...	&	&	...	&	...	&	...	\\
1999gp	&	LOWZ	&	0.0260 	&	16.029 	&	0.024 	&	1.705 	&	0.038 	&	0.008 	&	0.021 	&	&	...	&	...	&	...	&	...	&	...	&	...	&	Sb	&	...	&	...	&	...	&	&	...	&	...	&	...	\\
2000bh	&	LOWZ	&	0.0242 	&	15.937 	&	0.034 	&	-0.136 	&	0.073 	&	0.006 	&	0.026 	&	&	35.276 	&	0.114 	&	-0.065 	&	0.059 	&	0.159 	&	0.061 	&	S	&	...	&	...	&	...	&	&	...	&	...	&	...	\\
2000ca	&	LOWZ	&	0.0245 	&	15.561 	&	0.025 	&	0.534 	&	0.071 	&	-0.115 	&	0.021 	&	&	35.340 	&	0.055 	&	-0.154 	&	0.057 	&	-0.089 	&	0.047 	&	Sbc	&	10.04 	&	0.34 	&	0.23 	&	&	-9.00 	&	0.35 	&	0.51 	\\
2000cf	&	LOWZ	&	0.0365 	&	17.051 	&	0.029 	&	-0.512 	&	0.084 	&	-0.040 	&	0.023 	&	&	36.367 	&	0.066 	&	-0.001 	&	0.071 	&	0.112 	&	0.061 	&	Sbc*	&	...	&	...	&	...	&	&	...	&	...	&	...	\\
2000cn	&	LOWZ	&	0.0232 	&	16.549 	&	0.027 	&	-2.485 	&	0.149 	&	0.085 	&	0.022 	&	&	...	&	...	&	...	&	...	&	...	&	...	&	Scd	&	...	&	...	&	...	&	&	...	&	...	&	...	\\
2000dk	&	LOWZ	&	0.0164 	&	15.364 	&	0.027 	&	-2.047 	&	0.077 	&	-0.023 	&	0.022 	&	&	...	&	...	&	...	&	...	&	...	&	...	&	E	&	11.54 	&	1.37 	&	0.02 	&	&	-11.24 	&	0.56 	&	1.61 	\\
2000fa	&	LOWZ	&	0.0218 	&	15.889 	&	0.028 	&	0.490 	&	0.082 	&	0.043 	&	0.022 	&	&	35.000 	&	0.085 	&	-0.137 	&	0.074 	&	0.368 	&	0.064 	&	Im	&	9.82 	&	0.17 	&	0.28 	&	&	-8.65 	&	0.46 	&	0.47 	\\
2001V	&	LOWZ	&	0.0583 	&	17.632 	&	0.055 	&	1.354 	&	0.622 	&	-0.052 	&	0.030 	&	&	...	&	0.109 	&	-0.230 	&	0.060 	&	-0.040 	&	0.077 	&	Sbc	&	...	&	...	&	...	&	&	...	&	...	&	...	\\
2001ah	&	LOWZ	&	0.0406 	&	16.923 	&	0.029 	&	0.903 	&	0.233 	&	-0.084 	&	0.025 	&	&	37.355 	&	0.084 	&	0.003 	&	0.113 	&	-0.032 	&	0.083 	&	S	&	10.68 	&	0.13 	&	0.14 	&	&	-9.38 	&	1.02 	&	0.51 	\\
2001az	&	LOWZ	&	0.0305 	&	16.192 	&	0.031 	&	0.197 	&	0.093 	&	-0.170 	&	0.039 	&	&	36.523 	&	0.053 	&	-0.087 	&	0.047 	&	-0.145 	&	0.041 	&	Sbc	&	10.98 	&	0.50 	&	0.43 	&	&	-8.94 	&	1.38 	&	0.77 	\\
2001ba	&	LOWZ	&	0.0153 	&	14.717 	&	0.030 	&	0.459 	&	0.084 	&	-0.032 	&	0.022 	&	&	36.004 	&	0.066 	&	-0.289 	&	0.073 	&	0.325 	&	0.052 	&	...	&	...	&	...	&	...	&	&	...	&	...	&	...	\\
2001bf	&	LOWZ	&	0.0144 	&	15.281 	&	0.032 	&	-0.906 	&	0.052 	&	0.157 	&	0.026 	&	&	34.037 	&	...	&	...	&	...	&	...	&	...	&	Sbc	&	...	&	...	&	...	&	&	...	&	...	&	...	\\
2001bt	&	LOWZ	&	0.0154 	&	15.277 	&	0.030 	&	-0.537 	&	0.040 	&	0.123 	&	0.022 	&	&	...	&	0.056 	&	0.026 	&	0.050 	&	0.381 	&	0.052 	&	Sc	&	...	&	...	&	...	&	&	...	&	...	&	...	\\
2001cn	&	LOWZ	&	0.0163 	&	15.055 	&	0.031 	&	0.146 	&	0.061 	&	0.049 	&	0.026 	&	&	34.123 	&	0.067 	&	-0.107 	&	0.054 	&	0.288 	&	0.053 	&	Sc	&	...	&	...	&	...	&	&	...	&	...	&	...	\\
2001cz	&	LOWZ	&	0.0363 	&	16.604 	&	0.026 	&	1.582 	&	0.166 	&	-0.044 	&	0.023 	&	&	34.261 	&	0.058 	&	-0.351 	&	0.048 	&	0.121 	&	0.050 	&	Sb	&	...	&	...	&	...	&	&	...	&	...	&	...	\\
2001eh	&	LOWZ	&	0.0155 	&	15.111 	&	0.030 	&	-1.071 	&	0.054 	&	0.018 	&	0.024 	&	&	36.248 	&	0.081 	&	0.128 	&	0.075 	&	0.122 	&	0.068 	&	Sbc*	&	10.38 	&	0.15 	&	0.15 	&	&	-10.45 	&	0.26 	&	0.89 	\\
2001en	&	LOWZ	&	0.0129 	&	14.887 	&	0.030 	&	-1.026 	&	0.061 	&	0.056 	&	0.021 	&	&	34.248 	&	0.056 	&	0.140 	&	0.083 	&	0.328 	&	0.059 	&	Sb	&	10.37 	&	0.25 	&	0.04 	&	&	-10.22 	&	0.24 	&	0.98 	\\
2001ep	&	LOWZ	&	0.0144 	&	14.657 	&	0.030 	&	0.634 	&	0.094 	&	-0.056 	&	0.023 	&	&	33.832 	&	0.083 	&	-0.156 	&	0.069 	&	0.129 	&	0.059 	&	Sa	&	10.22 	&	0.11 	&	0.12 	&	&	-9.72 	&	0.87 	&	0.63 	\\
2001fe	&	LOWZ	&	0.0312 	&	16.666 	&	0.042 	&	-0.588 	&	0.119 	&	-0.024 	&	0.027 	&	&	34.155 	&	0.122 	&	0.039 	&	0.114 	&	0.327 	&	0.097 	&	E	&	10.99 	&	0.07 	&	0.21 	&	&	-10.88 	&	1.12 	&	0.20 	\\
2001ie	&	LOWZ	&	0.0162 	&	14.556 	&	0.026 	&	0.917 	&	0.042 	&	-0.050 	&	0.020 	&	&	35.732 	&	...	&	...	&	...	&	...	&	...	&	Sb	&	10.78 	&	0.09 	&	0.19 	&	&	-10.43 	&	0.50 	&	0.35 	\\
2002G	&	LOWZ	&	0.0247 	&	16.326 	&	0.047 	&	-0.307 	&	0.131 	&	0.144 	&	0.031 	&	&	...	&	0.096 	&	-0.182 	&	0.074 	&	0.449 	&	0.088 	&	Sb	&	10.62 	&	0.07 	&	0.12 	&	&	-10.49 	&	0.34 	&	0.32 	\\
2002bf	&	LOWZ	&	0.0302 	&	16.295 	&	0.041 	&	0.101 	&	0.089 	&	-0.069 	&	0.024 	&	&	35.355 	&	0.093 	&	-0.084 	&	0.079 	&	0.084 	&	0.068 	&	Sb	&	10.97 	&	0.08 	&	0.22 	&	&	-10.16 	&	0.78 	&	0.69 	\\
2002ck	&	LOWZ	&	0.0103 	&	14.201 	&	0.034 	&	-0.495 	&	0.062 	&	-0.037 	&	0.022 	&	&	35.826 	&	0.074 	&	0.027 	&	0.072 	&	0.192 	&	0.059 	&	Scd	&	9.48 	&	0.09 	&	0.29 	&	&	-8.65 	&	0.39 	&	0.40 	\\
2002cr	&	LOWZ	&	0.0281 	&	16.654 	&	0.027 	&	0.422 	&	0.359 	&	0.092 	&	0.023 	&	&	33.457 	&	0.073 	&	-0.066 	&	0.175 	&	0.440 	&	0.110 	&	S	&	10.83 	&	0.12 	&	0.03 	&	&	-9.68 	&	0.22 	&	0.17 	\\
2002de	&	LOWZ	&	0.0104 	&	13.955 	&	0.033 	&	0.006 	&	0.129 	&	0.063 	&	0.023 	&	&	35.626 	&	0.075 	&	-0.161 	&	0.143 	&	0.352 	&	0.069 	&	E	&	11.34 	&	0.13 	&	0.03 	&	&	-10.76 	&	0.48 	&	0.16 	\\
2002dj	&	LOWZ	&	0.0105 	&	14.570 	&	0.034 	&	-0.317 	&	0.209 	&	0.063 	&	0.023 	&	&	33.152 	&	0.062 	&	-0.065 	&	0.130 	&	0.426 	&	0.082 	&	Sc	&	10.40 	&	0.36 	&	0.35 	&	&	-9.38 	&	0.99 	&	0.57 	\\
2002dp	&	LOWZ	&	0.0346 	&	17.584 	&	0.058 	&	-1.538 	&	0.390 	&	0.208 	&	0.041 	&	&	33.584 	&	...	&	...	&	...	&	...	&	...	&	E	&	...	&	...	&	...	&	&	...	&	...	&	...	\\
2002ha	&	LOWZ	&	0.0134 	&	14.702 	&	0.030 	&	-1.354 	&	0.073 	&	-0.086 	&	0.023 	&	&	34.040 	&	0.067 	&	0.216 	&	0.104 	&	-0.007 	&	0.081 	&	Sab	&	11.09 	&	0.13 	&	0.14 	&	&	-10.43 	&	0.36 	&	0.76 	\\
2002he	&	LOWZ	&	0.0248 	&	16.257 	&	0.037 	&	-1.756 	&	0.168 	&	-0.035 	&	0.026 	&	&	...	&	...	&	...	&	...	&	...	&	...	&	E	&	11.12 	&	0.48 	&	0.90 	&	&	-8.61 	&	3.39 	&	0.57 	\\
2002hu	&	LOWZ	&	0.0382 	&	16.613 	&	0.024 	&	0.320 	&	0.095 	&	-0.110 	&	0.022 	&	&	36.297 	&	0.058 	&	-0.218 	&	0.054 	&	0.043 	&	0.048 	&	S	&	10.27 	&	1.44 	&	0.66 	&	&	-9.68 	&	1.27 	&	1.68 	\\
2002jy	&	LOWZ	&	0.0187 	&	15.730 	&	0.029 	&	0.865 	&	0.107 	&	-0.039 	&	0.022 	&	&	35.217 	&	0.062 	&	-0.221 	&	0.067 	&	0.167 	&	0.057 	&	Sc	&	10.46 	&	0.14 	&	0.11 	&	&	-9.43 	&	0.46 	&	0.48 	\\
2002kf	&	LOWZ	&	0.0195 	&	15.666 	&	0.030 	&	-1.111 	&	0.078 	&	-0.053 	&	0.023 	&	&	35.059 	&	0.070 	&	0.195 	&	0.066 	&	-0.100 	&	0.063 	&	...	&	...	&	...	&	...	&	&	...	&	...	&	...	\\
2003U	&	LOWZ	&	0.0279 	&	16.482 	&	0.033 	&	-2.156 	&	0.315 	&	-0.031 	&	0.028 	&	&	35.558 	&	0.081 	&	0.433 	&	0.131 	&	0.039 	&	0.113 	&	Scd	&	10.74 	&	0.30 	&	0.13 	&	&	-9.93 	&	0.33 	&	0.91 	\\
2003W	&	LOWZ	&	0.0211 	&	15.877 	&	0.026 	&	-0.091 	&	0.065 	&	0.111 	&	0.021 	&	&	34.828 	&	0.055 	&	-0.042 	&	0.051 	&	0.451 	&	0.041 	&	Sc	&	10.55 	&	0.40 	&	0.25 	&	&	-9.44 	&	1.18 	&	1.10 	\\
2003ch	&	LOWZ	&	0.0256 	&	16.675 	&	0.028 	&	-1.347 	&	0.155 	&	-0.039 	&	0.023 	&	&	36.019 	&	0.075 	&	0.169 	&	0.112 	&	-0.015 	&	0.085 	&	S0	&	...	&	...	&	...	&	&	...	&	...	&	...	\\
2003cq	&	LOWZ	&	0.0337 	&	17.194 	&	0.052 	&	-0.797 	&	0.165 	&	0.118 	&	0.044 	&	&	...	&	...	&	...	&	...	&	...	&	...	&	Sbc	&	...	&	...	&	...	&	&	...	&	...	&	...	\\
2003fa	&	LOWZ	&	0.0391 	&	16.679 	&	0.025 	&	1.474 	&	0.098 	&	-0.071 	&	0.022 	&	&	36.401 	&	0.048 	&	-0.318 	&	0.045 	&	0.022 	&	0.042 	&	S	&	10.81 	&	0.78 	&	0.61 	&	&	-9.57 	&	1.46 	&	1.13 	\\
2003ic	&	LOWZ	&	0.0542 	&	17.607 	&	0.030 	&	-1.999 	&	0.203 	&	-0.066 	&	0.029 	&	&	...	&	...	&	...	&	...	&	...	&	...	&	S0	&	11.70 	&	0.03 	&	0.27 	&	&	-10.76 	&	1.24 	&	0.07 	\\
2003it	&	LOWZ	&	0.0240 	&	16.353 	&	0.032 	&	-1.648 	&	0.168 	&	0.021 	&	0.029 	&	&	35.315 	&	0.098 	&	0.410 	&	0.140 	&	0.106 	&	0.110 	&	S	&	...	&	...	&	...	&	&	...	&	...	&	...	\\
2003iv	&	LOWZ	&	0.0335 	&	16.973 	&	0.028 	&	-2.157 	&	0.188 	&	-0.095 	&	0.027 	&	&	36.354 	&	0.092 	&	0.596 	&	0.159 	&	-0.362 	&	0.133 	&	E*	&	...	&	...	&	...	&	&	...	&	...	&	...	\\
2003kc	&	LOWZ	&	0.0343 	&	17.157 	&	0.033 	&	-0.660 	&	0.181 	&	0.110 	&	0.017 	&	&	...	&	...	&	...	&	...	&	...	&	...	&	Sc	&	10.67 	&	0.14 	&	0.09 	&	&	-9.38 	&	0.61 	&	0.31 	\\
2004L	&	LOWZ	&	0.0334 	&	17.391 	&	0.039 	&	-1.165 	&	0.240 	&	0.188 	&	0.031 	&	&	...	&	...	&	...	&	...	&	...	&	...	&	S	&	10.35 	&	0.15 	&	0.21 	&	&	-9.90 	&	0.59 	&	1.35 	\\
2004as	&	LOWZ	&	0.0321 	&	16.980 	&	0.028 	&	0.294 	&	0.142 	&	0.058 	&	0.024 	&	&	36.171 	&	0.069 	&	-0.199 	&	0.064 	&	0.381 	&	0.061 	&	Irr	&	9.28 	&	0.13 	&	0.08 	&	&	-9.23 	&	0.35 	&	0.44 	\\
2004bg	&	LOWZ	&	0.0219 	&	15.628 	&	0.044 	&	0.509 	&	0.119 	&	-0.043 	&	0.029 	&	&	35.217 	&	0.089 	&	-0.175 	&	0.069 	&	0.014 	&	0.106 	&	Scd	&	...	&	...	&	...	&	&	...	&	...	&	...	\\
\bottomrule
\end{tabular}
\end{adjustbox}
\tabnote{The entire catalog is available from the corresponding author, Y.-L.Kim, upon request.}
\end{center}
\end{table*}

In order to look for the systematic trends for both fitters and SN samples we employed, we analyze the output of the YONSEI SN Catalog. From this analysis, we would provide light-curve fit parameter areas for the study of the SN cosmology (the cosmology cut) and also examine the consistency between fitters as pointed out by previous studies.

\subsubsection{Light-Curve Shape and Color or Extinction Values as a Function of Redshift\label{sec.2.3.1}}

\begin{figure}[t]
\centering
\includegraphics[angle=-90,width=\columnwidth]{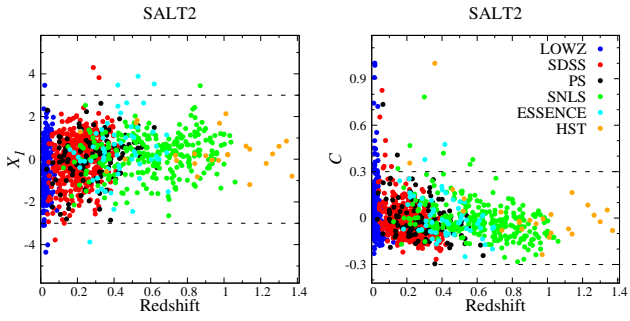}
\caption[Distribution of SALT2 fit parameters in the YONSEI All sample: $X_{1}$ and $C$ versus redshift]{Distribution of SALT2 fit parameters in the YONSEI All sample: $X_{1}$ (left panel) and $C$ (right panel) versus redshift. Faint (low $X_{1}$) and red (high $C$) SNe Ia are not found at the higher redshift range. SNe are colored by surveys. The dashed lines indicate our cosmology cut criteria we employed for SALT2 ($-$3 $<$ $X_{1}$ $<$ 3 and $-$0.3 $<$ $C$ $<$ 0.3).}
\label{fig.2.5}
\end{figure}

\begin{figure}[t]
\centering
\includegraphics[angle=-90,width=\columnwidth]{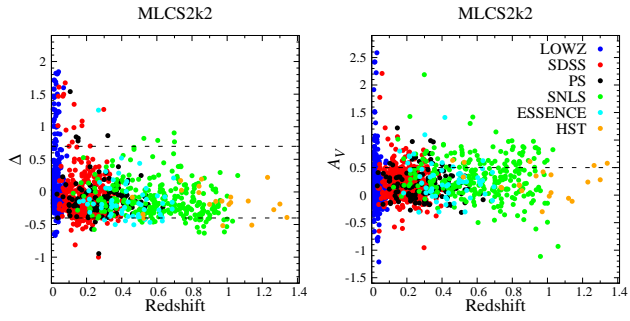}
\caption[Distribution of MLCS2k2 fit parameters in the YONSEI All sample: $\Delta$ and $A_{V}$ vs. redshift]{Same as Figure~\ref{fig.2.5}, but for MLCS2k2 fit parameters: $\Delta$ (left panel) and $A_{V}$ (right panel). Faint (high $\Delta$) and highly reddened (high $A_{V}$) SNe Ia are not found at the higher redshift range. The dashed lines indicate our cosmology cut criteria we employed for MLCS2k2 ($-$0.4 $<$ $\Delta$ $<$ 0.7 and $A_{V}$ $<$ 0.5).}
\label{fig.2.6}
\end{figure}

In Figures~\ref{fig.2.5} and \ref{fig.2.6}, we show the distributions of fitted shape and color or host extinction values of SALT2 and MLCS2k2 as a function of redshift, respectively. At the higher redshift range, both faint and red or highly reddened SNe Ia are not found. However, it is unclear whether this is because of the magnitude-limited surveys or the luminosity evolution of SNe Ia with redshift. We note that since we select the flat prior allowing negative $A_{V}$ values for our MLCS2k2 extinction distribution prior (see Section~\ref{sec.2.1}), there are many SNe with $A_{V} < 0$ on the right panel in Figure~\ref{fig.2.6}. However, the negative-$A_{V}$ distribution has little effect on the $\Delta$ distribution \citep{Kessler2009a}.

Considering the SNe Ia at the higher redshift range and the criteria adopted in \citet{Betoule2014} and \citet{Hicken2009b}, here we preliminarily define one of our cosmology cut criteria (see Section~\ref{sec.2.4.1}). For SALT2, we employ $-3 < X_{1} < 3$ and $-0.3 < C < 0.3$, and for MLCS2k2 we select $-0.4 < \Delta < 0.7$ and $A_{V} < 0.5$. As shown in Figures~\ref{fig.2.5} and \ref{fig.2.6}, most of SNe at the higher redshift range are included, except highly reddened ones in MLCS2k2.

\subsubsection{Comparison between SALT2 and MLCS2k2 Light-Curve Fit Parameters\label{sec.2.3.2}}

\begin{figure}[t]
\centering
\includegraphics[angle=0,width=60mm]{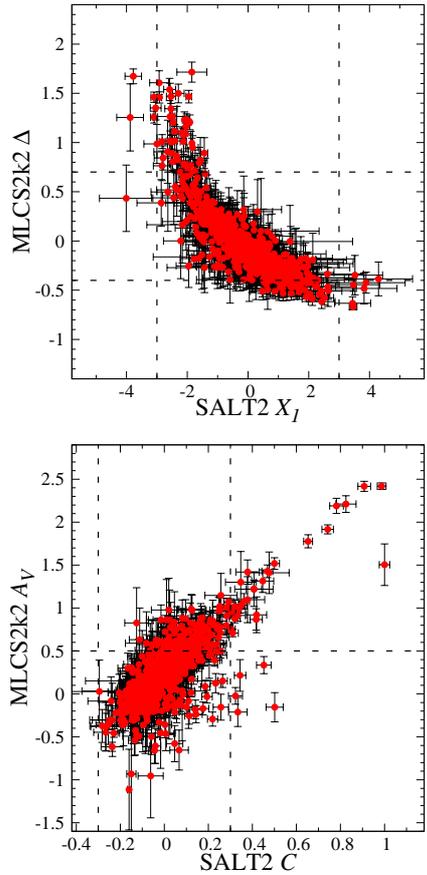}
\caption[Comparison between SALT2 and MLCS2k2 light-curve fit parameters]{Comparison between SALT2 and MLCS2k2 light-curve fit parameters. The upper panel shows the comparison of light-curve shape parameters, SALT2 $X_{1}$ versus MLCS2k2 $\Delta$, and the reddening parameters, SALT2 $C$ versus MLCS2k2 $A_{V}$, are compared in the lower panel. There is a good correlation between each other, as presented by \citet{Hicken2009b} and \citet{Sako2018}. The dashed lines indicate our cosmology cut criteria for both fitters, and many outliers are discarded from these criteria.}
\label{fig.2.7}
\end{figure}

Next, we compare the light-curve fit parameters between SALT2 and MLCS2k2 to examine the consistency with each other. In Figure~\ref{fig.2.7}, SALT2 $X_{1}$ is compared to MLCS2k2 $\Delta$ in the upper panel, and SALT2 $C$ is compared to MLCS2k2 $A_{V}$ in the lower panel. There is a good correlation between SALT2 and MLCS2k2 fit parameters, with a scatter and some outliers. $X_{1}$ and $\Delta$ are nonlinearly correlated, while $C$ and $A_{V}$ are linearly correlated, as presented by \citet{Hicken2009b} and \citet{Sako2018}. As shape parameters show nonlinearity, $\Delta$ spans a wide range in the vicinity of $X_{1} = -2$ (fast-declining), and the brightest SNe (negative $\Delta$) also have a wide range of $X_{1}$. Our cosmology cut criteria indicated by dashed lines exclude most of SNe in this nonlinear region. We note that there is a zeropoint offset for the color between the fitters, $C \approx -0.1$ when $A_{V} = 0$. Overall, light-curve fit parameters of SALT2 and MLCS2k2 we obtained show a good agreement between the two.

\subsubsection{Distribution of Shape and Color or Extinction Values for SALT2 and MLCS2k2\label{sec.2.3.3}}

\begin{figure}[t]
\centering
\includegraphics[angle=0,width=60mm]{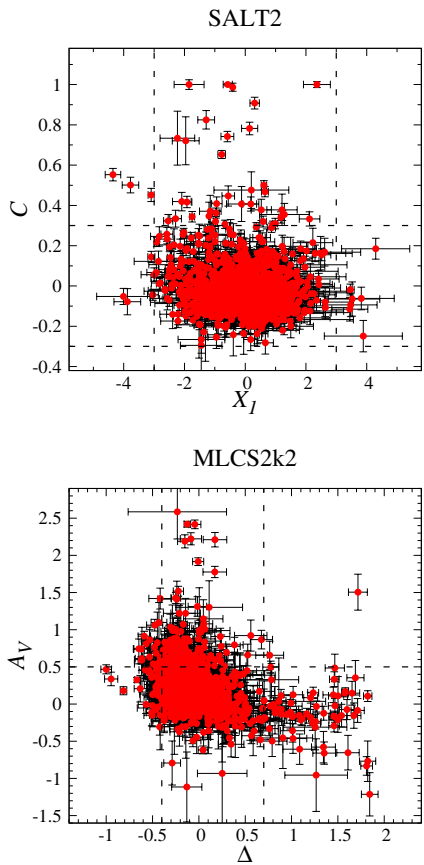}
\caption[Distribution of shape and color/extinction values for SALT2 and MLCS2k2]{Distribution of shape and color or host extinction values for SALT2 (upper panel) and MLCS2k2 (lower panel). We expect that peculiar SNe Ia are mostly placed in the scattered region. The dashed lines indicate our cosmology cut criteria, and most of peculiar SNe are removed by these cuts.}
\label{fig.2.8}
\end{figure}

Finally, we show the distribution of shape and color or host extinction values for SALT2 and MLCS2k2 in Figure~\ref{fig.2.8} to investigate the systematic trends in the YONSEI All sample. From the definition of the fitters, SNe Ia form a well-defined cluster of points centered on zero in the parameter space \citep[e.g.,][]{Campbell2013}. We expect that peculiar SNe Ia are mostly placed in the scattered region. As shown in the figure, our cosmology cut criteria we employed discard most of the scattered (i.e., peculiar) SNe Ia.

\subsection{YONSEI Hubble--Lem\^aitre Diagram\label{sec.2.4}}

\subsubsection{YONSEI Cosmology Sample\label{sec.2.4.1}}

In order to estimate the best-fit cosmological parameters, we should consider selecting only well-fitted normal SNe Ia from the YONSEI All sample compiled above. For this (the YONSEI `Cosmology' sample), we require the cosmology cut criteria. Following the criteria investigated in the previous section and adopted in \citet{Betoule2014} and \citet{Hicken2009b}, the cuts are based on the 1) the Milky Way extinction, 2) the uncertainty on the time of maximum in $B$-band ($t_{0}$), 3) the uncertainty on the light-curve shape parameter, and 4) the fitted shape and color or extinction parameters of each light-curve fitter. We detail the criteria below and the number of cosmology cuts is listed in Table~\ref{tab.2.2}.
\newline

\underline{Cosmology cut criteria:}
\newline

1) $E(B-V)_{MW} < 0.15$ mag.

2) $\sigma(t_{0}) < 2$.

3) $\sigma(shape) < 1$.

4) $-3 < X_{1} < 3$ and $-0.3 < C < 0.3$ for SALT2,

\hspace{0.2cm} and $-0.4 < \Delta < 0.7$ and $A_{V} < 0.5$ for MLCS2k2.
\newline

We then use the JLA likelihood code \citep{Betoule2014} to estimate the best-fit cosmological parameters for the YONSEI Cosmology sample assuming the flat $\Lambda$CDM model. The best-fit cosmological parameters obtained for our main cosmology sample were $\Omega$$_{M}$ $=$ 0.30, $\alpha$ $=$ 0.15, $\beta$ $=$ 3.69, and $M_{B}$ $=$ -19.06 for SALT2, and $H_{0}$ $=$ 63 and $\Omega$$_{M}$ $=$ 0.43 for MLCS2k2\footnote{This is most likely due to the lack of recent calibration of MLCS2k2 by using the SNe Ia at the high-redshift range \citep[see e.g.,][]{Guy2010, Betoule2014, Jones2015}.}. During estimating cosmological parameters, we applied Chauvenet's criterion \citep{Taylor1997} to reject outliers, removing SNe whose probability of obtaining a deviation from a mean is less than 1/(2$\times$sample size), assuming a Gaussian distribution of intrinsic luminosities. For our samples, this corresponds to 3.5$\sigma$ for SALT2 and 3.4$\sigma$ for MLCS2k2. This criterion does not significantly depend on the choice of cosmological models, and therefore the same SNe are excluded for all choices. The number of cuts by Chauvenet's criterion is listed in Table~\ref{tab.2.2}. Finally, of the 1182 SNe, 1049 pass the cosmology cut and Chauvenet's requirement for SALT2, and of the 1188 SNe, 821 pass for MLCS2k2. 

\subsubsection{Hubble--Lem\^aitre Diagram for the YONSEI Cosmology Sample\label{sec.2.4.2}}

\begin{figure}[t]
\centering
\includegraphics[angle=-90,width=\columnwidth]{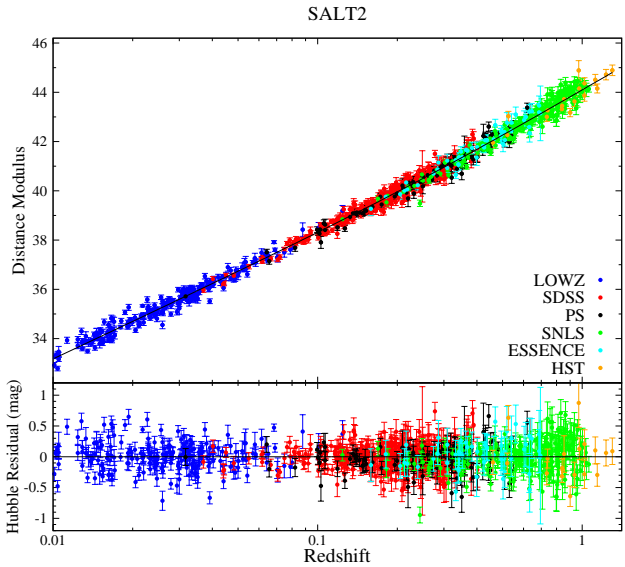}
\caption[YONSEI Hubble--Lem\^aitre diagram and residulas for the SALT2 sample]{YONSEI Hubble--Lem\^aitre diagram and residuals for the SALT2 sample. The solid line represents the best-fit flat $\Lambda$CDM cosmology parameters for SNe Ia alone. The values we obtained are $\Omega$$_{M}$ $=$ 0.30, $\alpha$ $=$ 0.15, $\beta$ $=$ 3.69, and $M_{B}$ $=$ -19.06. HRs from the best-fit are shown in the bottom panel. SNe are colored by surveys.}
\label{fig.2.9}
\end{figure}

\begin{figure}[t]
\centering
\includegraphics[angle=-90,width=\columnwidth]{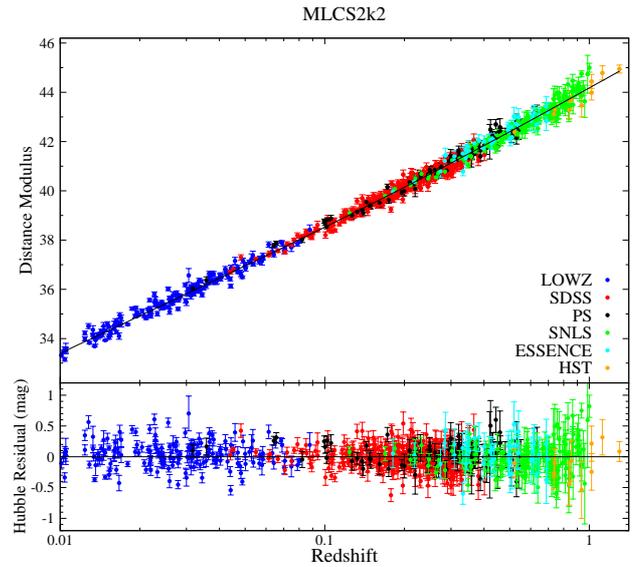}
\caption[YONSEI Hubble diagram and residuals for the MLCS2k2 sample]{Same as Figure~\ref{fig.2.9}, but for the MLCS2k2 sample. Best-fit flat $\Lambda$CDM cosmology parameters we obtained are $H_{0}$ $=$ 63 and $\Omega$$_{M}$ $=$ 0.43.}
\label{fig.2.10}
\end{figure}

From the YONSEI Cosmology sample and best-fit cosmological parameters we obtained above, finally, here we present the YONSEI Hubble--Lem\^aitre diagram in Figures~\ref{fig.2.9} and \ref{fig.2.10} for SALT2 and MLCS2k2, respectively. Hubble residuals (HRs $\equiv \mu_\mathrm{SN}-\mu_\mathrm{model}(z)$, where $\mu_\mathrm{model}(z)$ is the predicted distance modulus based on the cosmological model) from the best-fit are also presented in the bottom panel. 

Table~\ref{tab.2.3} shows the weighted means and rms scatters of the HRs for each data set. The average weighted mean of the YONSEI Cosmology sample is $0.00\pm0.01$ mag, and the average rms scatter is $\sim$0.19 mag. When we look at the table by data sets, the HST sample has a more negative weighted mean of HRs than other samples. This means that the HST sample is brighter, after light-curve corrections, than samples in other surveys. However, as discussed in Section~\ref{sec.2.3.1}, it is not clear whether this is due to the selection bias or intrinsic differences in SNe Ia.  Because of the larger photometric uncertainties, the HST and ESSENCE samples show larger values in the rms scatter. However, our rms values for both samples are similar to those in previous studies \citep[see the rms values for the high-z sample in][]{Hicken2009b, Kessler2009a, Conley2011, Suzuki2012}.

\begin{table*}
\centering
\caption{\label{tab.2.3} The Weighted Mean and rms Scatter of Hubble Residuals in Each Data Set}
\begin{tabular}{lccrcccccrccc}
\hline\hline
 & & \multicolumn{5}{c}{SALT2} & & \multicolumn{5}{c}{MLCS2k2} \\
\cline{3-7} \cline{9-13}
SN Data & & $N_{SN}$ & \multicolumn{1}{c}{HR$_{WM}$} & Error &  rms  & Error && $N_{SN}$ & HR$_{WM}$                 & Error &  rms  & Error \\
& &          & \multicolumn{1}{c}{(mag)}     & (mag) & (mag) & (mag) &&          & \multicolumn{1}{c}{(mag)} & (mag) & (mag) & (mag) \\
\hline
\textbf{YONSEI} & & 1049 & 0.000 & 0.006 & 0.194 & 0.004 && 821 & 0.006 & 0.006 & 0.182 & 0.004 \\
\hline
LOWZ    & & 218 &  0.020   &  0.015 &  0.206 & 0.010 && 174 &  0.011   & 0.015 &  0.188 & 0.010 \\
SDSS    & & 392 & $-0.006$ &  0.008 &  0.165 & 0.006 && 328 & $-0.003$ & 0.008 &  0.153 & 0.006 \\
PS      & & 108 & $-0.022$ &  0.017 &  0.172 & 0.012 && 98  &  0.056   & 0.016 &  0.156 & 0.011 \\
ESSENCE & & 51  &  0.016   &  0.033 &  0.235 & 0.023 && 41  &  0.049   & 0.036 &  0.230 & 0.026 \\
SNLS    & & 262 & $-0.006$ &  0.012 &  0.197 & 0.009 && 170 & $-0.014$ & 0.015 &  0.191 & 0.010 \\
HST     & & 18  & $-0.116$ &  0.069 &  0.294 & 0.051 && 10  & $-0.054$ & 0.072 &  0.227 & 0.053 \\
\hline
\end{tabular}
\end{table*}

\begin{figure}[t]
\centering
\includegraphics[angle=0,width=70mm]{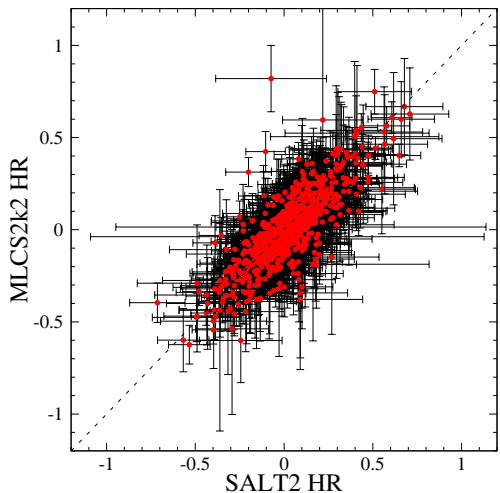}
\caption[Correlation between Hubble residuals obtained from SALT2 and MLCS2k2]{Correlation between HRs obtained from SALT2 and MLCS2k2. They show a good agreement with a mean offset of 0.02 mag, and the value of the correlation coefficient is 0.86. 718 SNe Ia are cross-matched on SALT2 and MLCS2k2. There is an outlier, 06D2cb in the SNLS sample, for which only one filter is used when MLCS2k2 estimates light-curve fit parameters. The dashed line is for the one-to-one relation.}
\label{fig.2.11}
\end{figure}

\begin{figure}[h]
\centering
\includegraphics[angle=0,width=65mm]{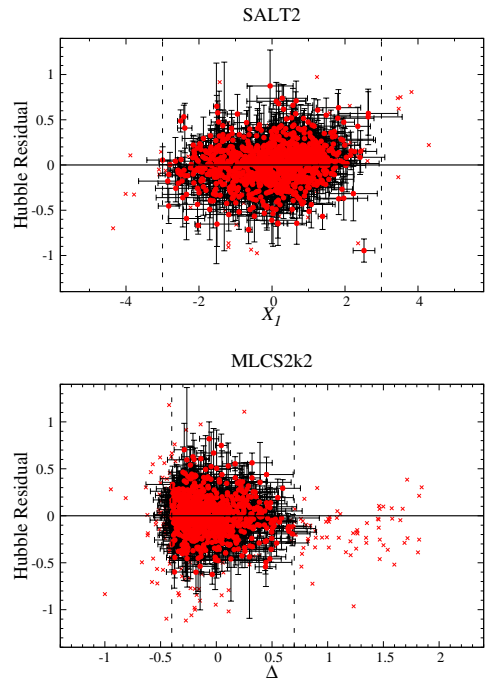}
\caption[Hubble residuals versus light-curve shape parameters for SALT2 and MLCS2k2]{HRs versus light-curve shape parameters for SALT2 ($X_{1}$; upper panel) and MLCS2k2 ($\Delta$; lower panel). No significant trends exist for both fitters. Outliers (mostly peculiar SNe Ia; cross marks) are appropriately removed by our cosmology cut criteria (dashed lines). Note a peculiar region ($\Delta$ $>$ 0.7) in the lower panel, where SNe have mostly negative HR values. These 1991bg-like SNe Ia are also well removed by our cosmology cut criteria.}
\label{fig.2.12}
\end{figure}

\subsubsection{Systematic Tests for Hubble Residuals\label{sec.2.4.3}}

We now turn to find out systematic trends in HRs, because these trends, if any, could give an effect on our results described in the following sections. In Figure~\ref{fig.2.11}, we first present a correlation between the HRs obtained from SALT2 and MLCS2k2 fitters. They show a good agreement with a mean offset of 0.02 mag and the correlation coefficient of 0.86. As we show that SALT2 and MLCS2k2 are similarly characterizing the SN Ia light-curves (see Section~\ref{sec.2.3.2}), most of the residual scatters might be due to the different methods they used when both fitters are determining the distance modulus.

Figure~\ref{fig.2.12} shows HRs as a function of the light-curve shape parameters for SALT2 and MLCS2k2, respectively. No significant trends are found for both fitters. Outliers (cross marks), which are expected to be peculiar SNe Ia, are appropriately removed by our cosmology cut criteria (dashed lines). We note a peculiar region in the lower panel. In the region of $\Delta$ $>$ 0.7 for MLCS2k2, the residuals have mostly negative values as pointed out by \citet{Hicken2009b}. They discussed that this ``dip" is most likely due to 1991-bg like SNe Ia, which are fainter than normal SNe Ia. However, this class of SNe is also well discarded by our cosmology cut criteria.

\begin{figure}[t]
\centering
\includegraphics[angle=0,width=65mm]{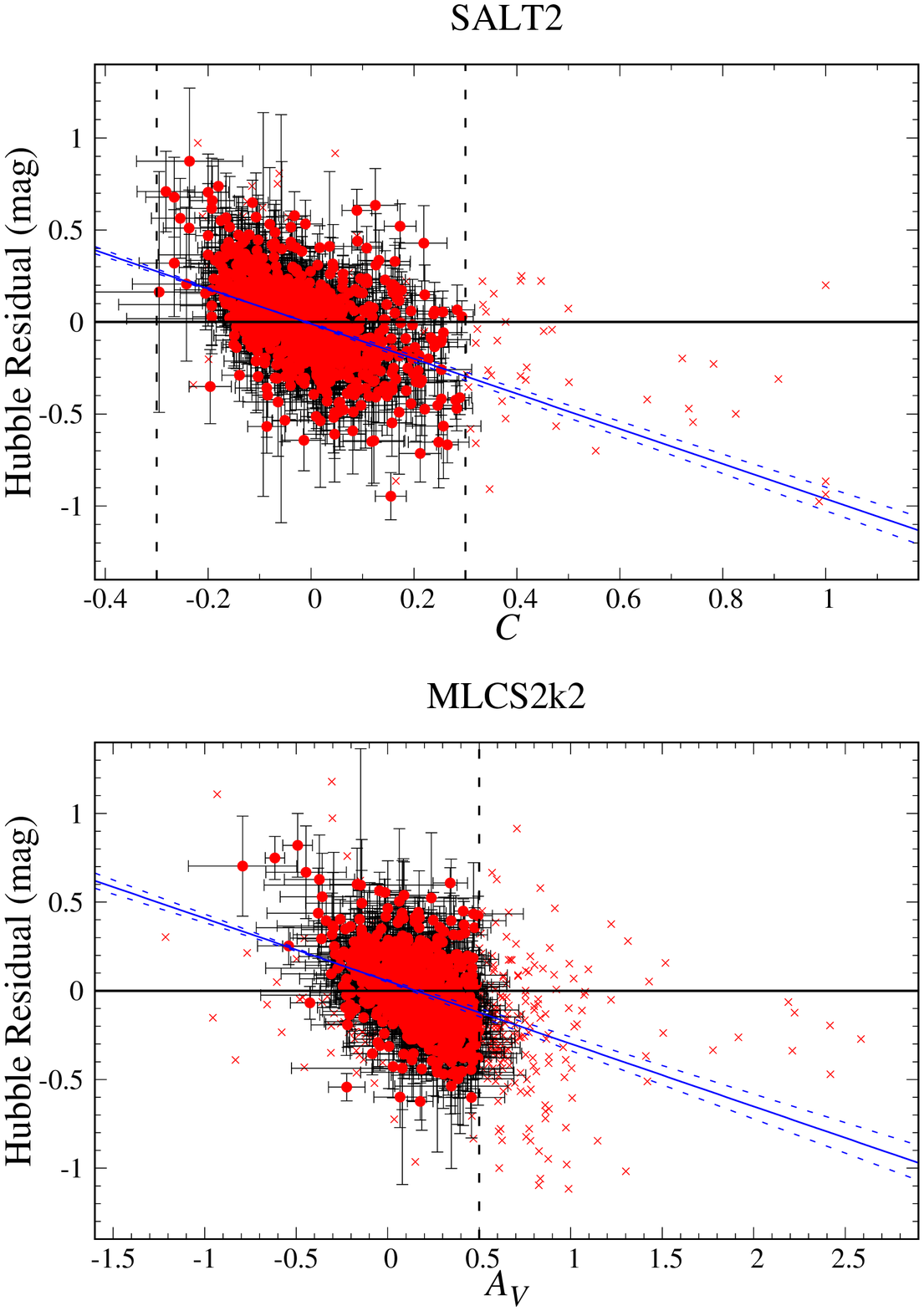}
\caption[Hubble residuals versus $C$ and $A_{V}$ for SALT2 and MLCS2k2]{HRs versus $C$ and $A_{V}$ for SALT2 (upper panel) and MLCS2k2 (lower panel). Most of SNe Ia are well-clustered, but there are negative trends in both fitters (blue lines with $\pm1\sigma$ ranges). The blaek dashed lines indicate our cosmology cut criteria. The redder ($C > 0.3$ for SALT2) and highly reddened ($A_{V} > 0.5$ for MLCS2k2) SNe, which have mostly negative residuals, are removed by these cut criteria.}
\label{fig.2.13}
\end{figure}

\begin{figure}[t]
\centering
\includegraphics[angle=0,width=60mm]{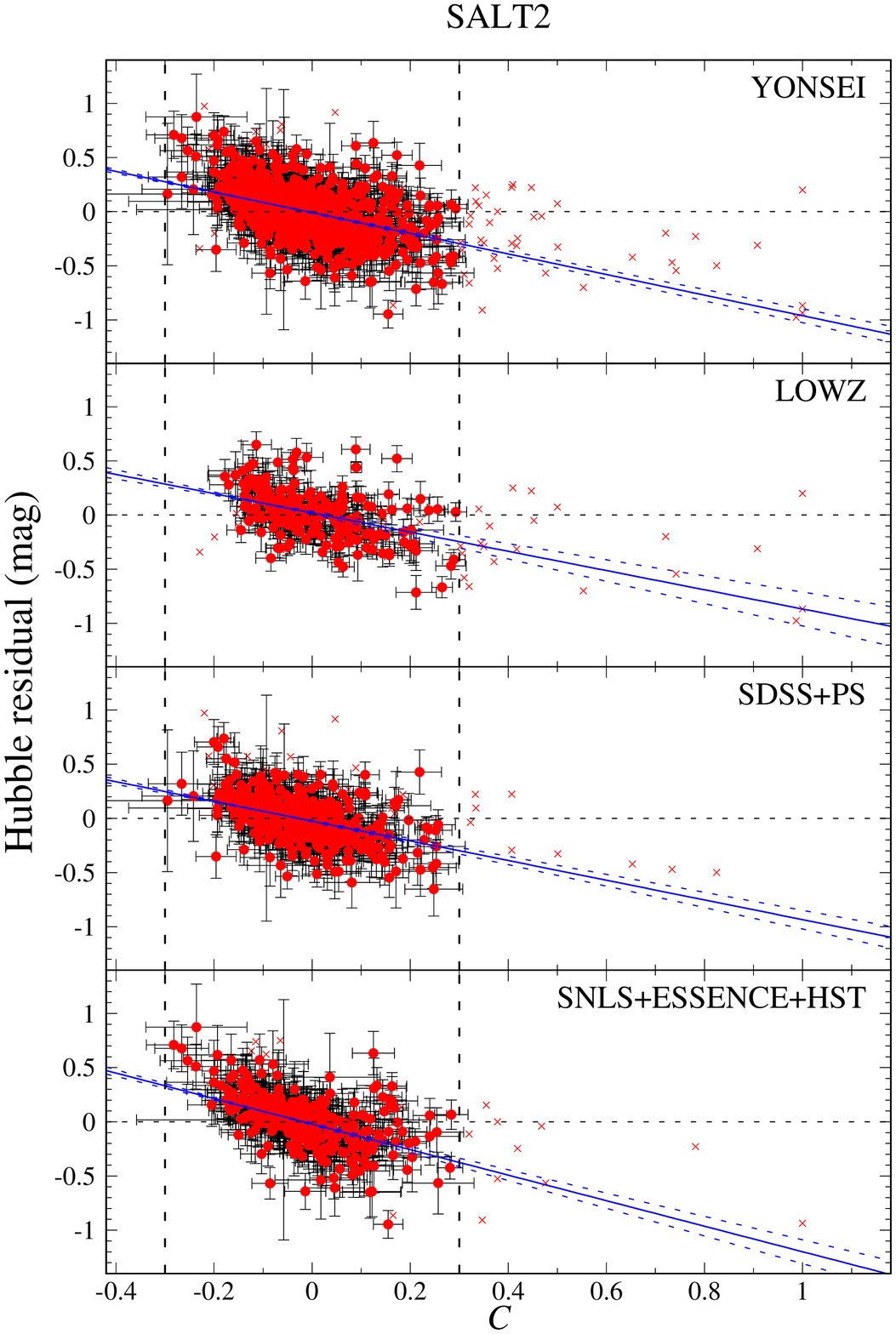}
\caption[SALT2 Hubble residuals versus SALT2 $C$ split by SN data]{SALT2 HRs versus SALT2 $C$ split by SN data. Negative trends are observed in every sample (blue lines with $\pm1\sigma$ ranges). }
\label{fig.2.14}
\end{figure}

\begin{figure}
\centering
\includegraphics[angle=0,width=60mm]{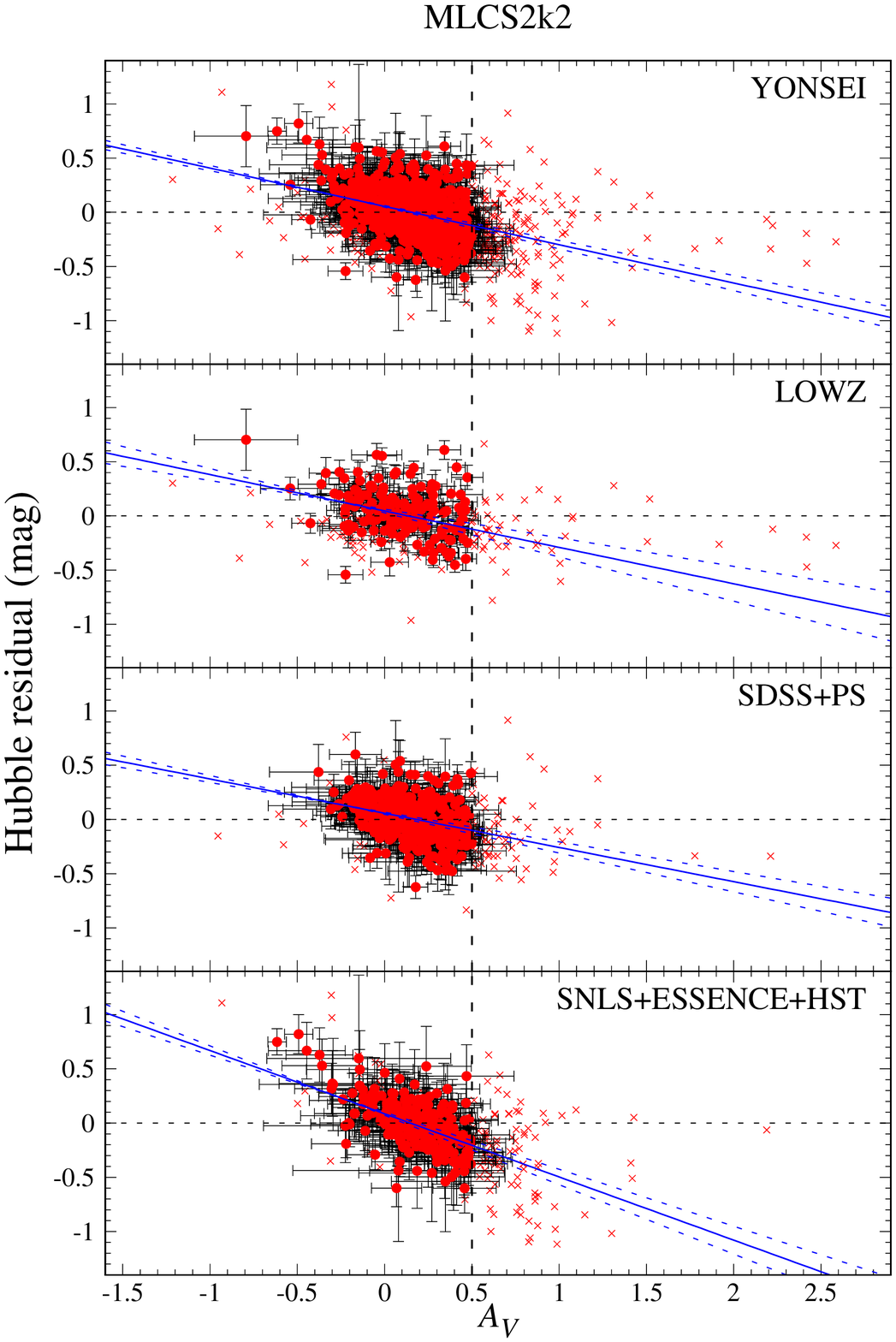}
\caption[MLCS2k2 Hubble residuals versus MLCS2k2 $A_{V}$ split by SN data]{Same as Figure~\ref{fig.2.14}, but for MLCS2k2 $A_{V}$. We observe a negative trend in every sample (blue lines with $\pm1\sigma$ ranges). }
\label{fig.2.15}
\end{figure}

Finally, we plot the HRs versus $C$ and $A_{V}$ for SALT2 and MLCS2k2 in Figure~\ref{fig.2.13}. At first sight, it looks like that there are negative trends between HRs and $C$ or $A_{V}$ for both fitters. Even though the redder ($C > 0.3$ for SALT2) and highly reddened ($A_{V} > 0.5$ for MLCS2k2) SNe, which have mostly negative residuals, are removed by our cosmology cut criteria, trends still remain. In order to investigate this, we have plotted the HRs versus $C$ and $A_{V}$ for SALT2 and MLCS2k2 with a sample split by redshift range in Figures~\ref{fig.2.14} and \ref{fig.2.15}, respectively. As shown in the figures, the negative trends are universal at all samples in the different redshift ranges.  However, \citet{Hicken2009b} showed that the cuts on $C$ and $A_{V}$ make this trend have little effect on the SN cosmology. On the other hand, we could consider that this trend would show a hint of the dependence of SN luminosities on host galaxies. For example, hosts which have highly reddened SNe might be related to star-forming environments (young populations), while those with little extinction values are related to passive environments (old populations).

\subsection{Host Galaxy Data\label{sec.2.5}}

Host galaxy properties, specifically morphological types, the stellar mass ($M_{stellar}$), and global specific star formation rate ($sSFR$; the star formation rate per unit stellar mass), are also required to explore the environmental dependence of SN Ia luminosities. Morphological types for the LOWZ sample (194 SNe out of 218) are drawn from the NASA Extragalactic Database (NED)\footnote{\url{http://ned.ipac.caltech.edu/}} and the HyperLeda database \citep{Makarov2014}\footnote{\url{http://leda.univ-lyon1.fr/}}, and from the Korea Institute for Advanced Science Value-Added Galaxy Catalog \citep[][KIAS-VAGC]{Choi2010}\footnote{\url{http://astro.kias.re.kr/vagc/dr7/}} and \citet{Han2010} for the SDSS sample (55 out of 392).

Host $M_{stellar}$ and global $sSFR$ are taken from \citet{Kim2018}, which provide the host data for LOWZ, SDSS, and SNLS samples\footnote{Since \citet{Kim2018} do not provide the host data for PS, ESSENCE, and HST samples, we do not include those samples in our host analysis.}. Briefly, they employ the P\'{E}GASE.2 galaxy spectral evolution code \citep{Fioc1997, LeBorgne2002, LeBorgne2004} to determine host galaxy properties, following the method described in detail in \citet{Sullivan2006, Sullivan2010}. They use a set of 14 exponentially declining star formation histories (SFHs) and foreground dust screens ranging from $E(B-V)=0$ to 0.30\,mag in steps of 0.05. Then, they fit the host galaxy data from \citet{Smith2012} and \citet{Sako2018} (for the SDSS sample), and \citet{Sullivan2010} (for the SNLS sample). Host galaxy properties for the LOWZ sample are taken from \citet{Neill2009} that uses the same P\'{E}GASE.2 approach. In total, for SALT2, 89 (out of 218) for LOWZ, 355 (out of 392) for SDSS, and 213 (out of 262) hosts for SNLS are matched with the YONSEI Cosmology sample, while for MLCS2k2, 73 (out of 174) for LOWZ, 292 (out of 328) for SDSS, and 147 (out of 170) hosts for SNLS are collected. The data and sample sizes used in our analysis are listed in Tables~\ref{tab.A.Catalog} and ~\ref{tab.2.4}, respectively. We note that as mentioned in \citet{Sullivan2010} and \citet{Kim2018}, we also restrict the SNLS sample to $z\le0.85$, where the SNLS sample has the highest signal-to-noise ratio and the Malmquist corrections are the smallest.

\begin{table*}
\centering
\caption{\label{tab.2.4} Summary of Host Data in the YONSEI SN Catalog}
\begin{adjustbox}{width=\textwidth,center=\textwidth}
\begin{tabular}{lcccccccccc}
\toprule
 & & \multicolumn{3}{c}{SALT2} & & \multicolumn{3}{c}{MLCS2k2} \\
\cline{3-6} \cline{8-11}
SN Data & & $N_{SN}$ & Morphology & Mass \& sSFR & Local           && $N_{SN}$ & Morphology & Mass \& sSFR & Local \\
 &  &               &                   &                        & Environment &&                &                   &                        & Environment \\
\midrule
LOWZ           &  & 218 & 194 &  89 &  40  && 174 & 152 &  73 & 29 \\
SDSS           &  & 392 &  55  & 355 & 203 && 328 &  45 & 292 & 164 \\
SNLS           &  & 262 &   0  & 213 & 130 && 170 &   0  & 147 & 96 \\
\hline
\textbf{YONSEI} & & 872 & 249 & 657 & 373 &  & 672 & 197 & 512 & 289 \\
\bottomrule
\end{tabular}
\end{adjustbox}
\end{table*}

\begin{figure}
\centering
\includegraphics[angle=0,width=60mm]{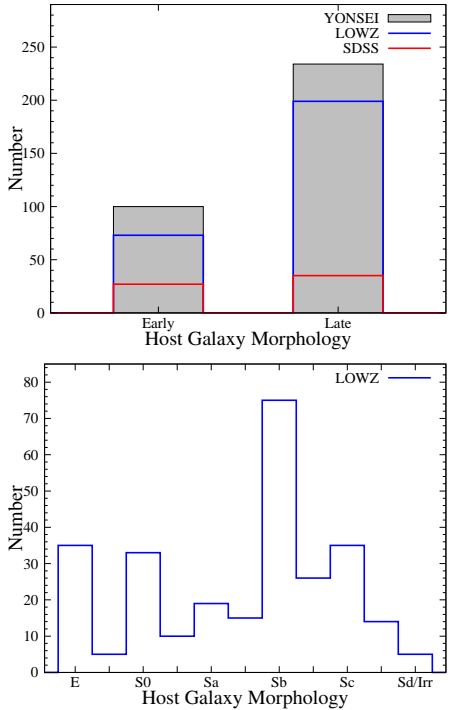}
\caption[Distribution of the YONSEI host sample in the host galaxy morphology]{Distribution of the YONSEI host sample in the host galaxy morphology. Only LOWZ (blue histogram) and SDSS (red histogram) samples have host morphological information (see text). The upper panel shows the distribution of those samples when we divide hosts as early- and late-type galaxies, and the lower panel is only for the LOWZ sample with more specific morphological types. SNe Ia prefer to explode in the late-type galaxies as expected.}
\label{fig.2.16}
\end{figure}

\begin{figure}
\centering
\includegraphics[angle=0,width=60mm]{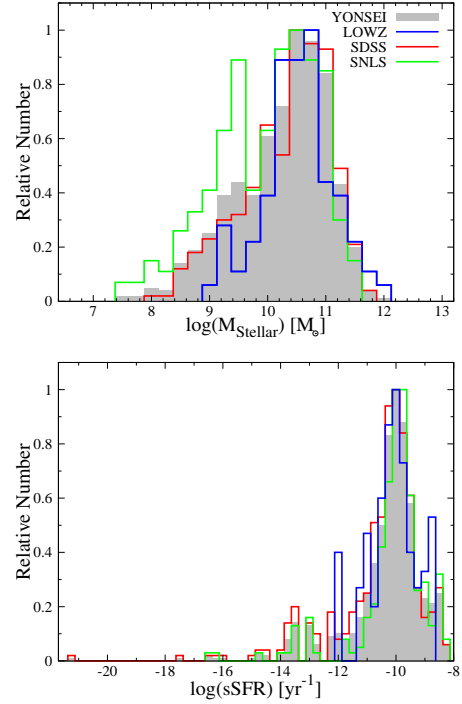}
\caption[Distribution of the YONSEI host sample in host stellar mass and $sSFR$]{Same as Figure~\ref{fig.2.16}, but in $M_{stellar}$ (upper panel) and $sSFR$ (lower panel). We note that the LOWZ sample (blue histogram) has more fraction of high-mass (log($M_{stellar}$) $> 10.0$) hosts than that in other samples. In $sSFR$, the distributions are fairly similar each other.}
\label{fig.2.17}
\end{figure}

Figures~\ref{fig.2.16} and \ref{fig.2.17} show the distributions of the YONSEI host sample in the host galaxy morphology, $M_{stellar}$, and $sSFR$ planes. As shown in the figures, SNe Ia prefer to explode in the late-type, star-forming, and more massive (high-mass; log($M_{stellar}$) $> 10.0$) systems. We note that the fraction of high-mass hosts in the LOWZ sample is much higher than that in other samples (see the blue histogram in the upper panel of Figure~\ref{fig.2.17}). This is because many SNe in the LOWZ sample were discovered in host-targeted surveys which prefer more luminous and massive galaxies, while the SDSS and SNLS samples were from untargeted rolling surveys without selection biases \citep[see also][]{Neill2009, Kelly2010, Sullivan2010, Pan2014}. In $sSFR$, the distributions of each sample (the lower panel of Figure~\ref{fig.2.17}) are fairly similar, as pointed out by \citet{Neill2009}.

\begin{figure}
\centering
\includegraphics[angle=-90,width=\columnwidth]{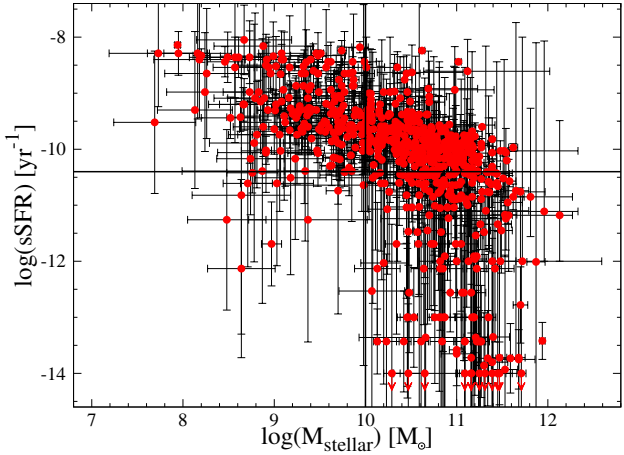}
\caption[Distribution of stellar mass and $sSFR$ for the YONSEI host sample]{Distribution of $M_{stellar}$ and $sSFR$ for the YONSEI host sample. As we expected, most of high-mass galaxies have the lower value of $sSFR$, while the low-mass galaxies universally show a strong star formation activity. The black horizontal line shows the cut criterion when we split galaxies into star-forming and passive environments at log($sSFR$) $= -10.4$. The black vertical line represents the cut criterion (log($M_{stellar}$) $= 10.0$) when we split host galaxies into high- and low-mass galaxies. }
\label{fig.2.18}
\end{figure}

We have also investigated the distribution of our host galaxies in the $M_{stellar}$--$sSFR$ plane in Figure~\ref{fig.2.18}. We see that most of the high-mass host galaxies have a lower value of $sSFR$, while the low-mass hosts universally show a strong star formation activity, as expected. 

Finally, we note that in the following sections, we will investigate the dependence of various SN Ia properties on their host galaxy properties. Therefore, we need to split the host galaxies into several different groups. The first group is split according to their morphology: early- and late-types. In the case of the LOWZ sample, we further split it more specifically as E--S0, S0a--Sc, and Scd/Sd/Irr galaxies, following the scheme presented in \citet{Hicken2009b}. The second group is split into high- and low-mass galaxies by a value of log($M_{stellar}$) $= 10.0$ \citep[the black vertical line in Figure~\ref{fig.2.18}; see][]{Sullivan2010, Childress2013, Pan2014}. The final group is split based on the log($sSFR$): passive and star-forming environments split at log($sSFR$) =$-10.4$ \citep[the black horizontal line in Figure~\ref{fig.2.18}; for the similar criteria, see][]{Sullivan2010, Childress2013, Rigault2013, Rigault2015, Pan2014, Jones2015}.

\subsubsection{\citet{Kim2018} Method to Infer the Local Environments of Type Ia Supernovae \label{sec.2.5.1}}

Recent studies of SN Ia host galaxy have focused on the local environments at the SN explosion site \citep[e.g.,][]{Rigault2013, Rigault2015, Rigault2018, Jones2015, Jones2018b}. However, the local environmental measurements are challenging, and so only SNe at the low-redshift range ($z<0.1$) were available for those studies. In order to select a sample over a wider redshift range, \citet{Kim2018} introduced an empirical method to infer the local environments, only based on the global properties of host galaxies, such as $M_{stellar}$ and global $sSFR$. The main idea is that SNe Ia in locally star-forming environments can be selected when their hosts are globally star-forming and low-mass galaxies. For the SNe Ia exploding in globally passive host galaxies, all of them are also in locally passive environments, which is demonstrated by \citet{Rigault2013}. We also apply this method to infer the local environments of our sample and the local sample sizes are also listed in Table~\ref{tab.2.4}.

\section{The Dependence of Type Ia Supernova Luminosities on the Host Galaxy Properties\protect\footnote{We note that in this paper we only present the result values related to the local environmental dependence of SN Ia properties. Figures and more detailed analysis of the local environmental study can be found in \citet{Kim2018}.}\label{sec:dependence}}

\subsection{Supernova Light-Curve Fit Parameters as a Function of Host Galaxy Properties\label{sec.3.2}}

In Figures~\ref{fig.S2.Fitparameters.Vs.Host} and \ref{fig.MLCS.Fitparameters.Vs.Host}, we plot the SN Ia light-curve fit parameters as a function of host global properties for SALT2 and MLCS2k2, respectively. The weighted mean of light-curve fit parameters in different host properties are summarized in Table~\ref{tab.Fitparameters.Vs.Host}.

\begin{table*}
\centering
\caption{\label{tab.Fitparameters.Vs.Host} The Weighted Mean of Light-Curve Fit Parameters in Different Host Properties}
\begin{adjustbox}{width=\textwidth,center=\textwidth}
\begin{tabular}{lccrcccrccccrcccrcc}
\toprule
Host     & Group & \multicolumn{8}{c}{SALT2} & & \multicolumn{8}{c}{MLCS2k2} \\
\cline{3-10} \cline{12-19}
Property &       & $N_{SN}$ & \multicolumn{1}{c}{$X_{1}$,$_{WM}$} && Error && \multicolumn{1}{c}{$C_{WM}$} && Error & & $N_{SN}$ & \multicolumn{1}{c}{$\Delta_{WM}$} && Error && \multicolumn{1}{c}{$A_{V, WM}$} && Error \\
\midrule
Morphology & E-S0       &  44 & $-1.515$ && 0.079 && $-0.011$ && 0.014 &&  32 &   0.213  && 0.040 && $-0.024$ && 0.029 \\
           & S0a-Sc     & 134 & $-0.238$ && 0.083 &&   0.007  && 0.009 && 106 & $-0.117$ && 0.019 &&   0.144  && 0.018 \\
           & Scd/Sd/Irr &  16 &   0.420  && 0.194 &&   0.013  && 0.018 &&  14 & $-0.163$ && 0.041 &&   0.167  && 0.035 \\
\hline
Diff.      & (Scd/Sd/Irr - E-S0) & & \textbf{1.935} && \textbf{0.209} && \textbf{0.024} && \textbf{0.023} && & \textbf{0.376} && \textbf{0.057} && \textbf{0.191} && \textbf{0.045} \\
\hline
\\
\hline
Morphology & Early-type &  68 & $-1.426$ && 0.079 && $-0.003$ && 0.011 &&  51 &   0.113  && 0.038 && 0.042 && 0.028 \\
           & Late-type  & 181 & $-0.180$ && 0.070 &&   0.011  && 0.007 && 146 & $-0.130$ && 0.015 && 0.159 && 0.015 \\
\hline
Diff.      & & & \textbf{1.246} && \textbf{0.106} && \textbf{0.014} && \textbf{0.013} && & \textbf{0.243} && \textbf{0.041} && \textbf{0.117} && \textbf{0.032}  \\
\hline
\\
\hline
Mass & High-mass & 469 & $-0.383$ && 0.046 && $-0.014$ && 0.004 && 369 & $-0.111$ && 0.011 && 0.168 && 0.009 \\
     & Low-mass  & 188 &   0.410  && 0.048 &&   0.016  && 0.007 && 143 & $-0.213$ && 0.011 && 0.166 && 0.013 \\
\hline
Diff. & & & \textbf{0.793} && \textbf{0.066} && \textbf{0.030} && \textbf{0.008} && & \textbf{0.102} && \textbf{0.016} && \textbf{0.002} && \textbf{0.016} \\
\hline
\\
\hline
sSFR & Globally Passive      & 196 & $-0.844$ && 0.070 && $-0.009$ && 0.007 && 155 & $-0.021$ && 0.020 && 0.140 && 0.015 \\
     & Globally Star-Forming & 461 & $-0.016$ && 0.041 && $-0.003$ && 0.004 && 357 & $-0.169$ && 0.011 && 0.175 && 0.009 \\
\hline
Diff. & & & \textbf{0.828} && \textbf{0.081} && \textbf{0.006} && \textbf{0.008} && & \textbf{0.148} && \textbf{0.023} && \textbf{0.035} && \textbf{0.017} \\
\hline
\\
\hline
sSFR & Locally Passive         & 196  & $-0.844$ && 0.070 && $-0.009$ && 0.007 && 155 & $-0.021$ && 0.020 && 0.140 && 0.015 \\
        & Locally Star-Forming  & 177  & 0.455      && 0.046 &&    0.017  && 0.007 && 134 & $-0.219$ && 0.011 && 0.167 && 0.013  \\
\hline
Diff.   &  &  & \textbf{1.299} && \textbf{0.084} && \textbf{0.026} && \textbf{0.010} && & \textbf{0.198} && \textbf{0.023} && \textbf{0.027} && \textbf{0.020} \\
\bottomrule
\end{tabular}
\end{adjustbox}
\end{table*}

\subsubsection{Light-Curve Shape: $X_{1}$ and $\Delta$\label{sec.LCShpae.Vs.Host}}

\begin{figure}[t]
\centering
\includegraphics[angle=-90,width=\columnwidth]{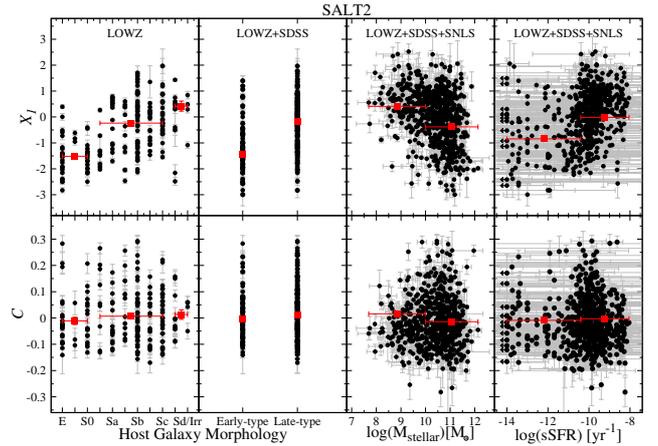}
\caption[SALT2 light-curve fit parameters as a function of host galaxy properties]{SALT2 light-curve fit parameters, $X1$ (top panels) and $C$ (bottom panels), as a function of host galaxy properties for the YONSEI host sample. The red squares represent the weighted means of each light-curve fit parameter in bins of host galaxy information.}
\label{fig.S2.Fitparameters.Vs.Host}
\end{figure}

\begin{figure}[t]
\centering
\includegraphics[angle=-90,width=\columnwidth]{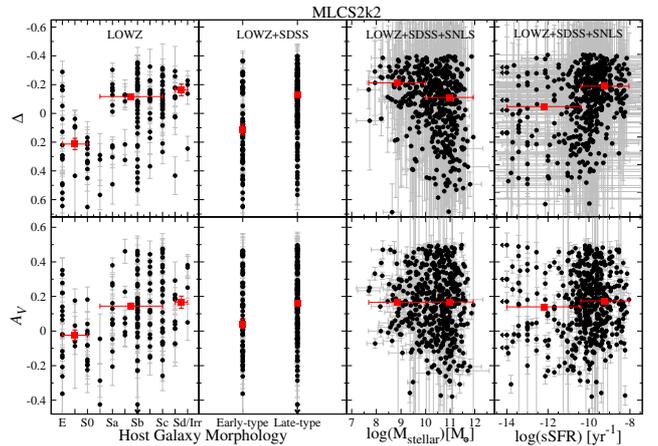}
\caption[MLCS2k2 light-curve fit parameters as a function of  host galaxy properties]{Sames as Figure~\ref{fig.S2.Fitparameters.Vs.Host}, but for MLCS2k2, $\Delta$ (top panels) and $A_{V}$ (bottom panels).}
\label{fig.MLCS.Fitparameters.Vs.Host}
\end{figure}

We start by examining the trend of SN Ia light-curve shape with host properties in the top panels of Figures~\ref{fig.S2.Fitparameters.Vs.Host} and \ref{fig.MLCS.Fitparameters.Vs.Host}. Overall, we recover the trend in previous works \citep[e.g.,][]{Filippenko1989, Hamuy1995, Hamuy1996a, Hamuy2000, Gallagher2005, Sullivan2006, Howell2009, Neill2009, Lampeitl2010, Smith2012, Childress2013, Pan2014}.

First, we see the light-curve shape trend with the host galaxy morphology in the first two panels of the top panels of Figures~\ref{fig.S2.Fitparameters.Vs.Host} and \ref{fig.MLCS.Fitparameters.Vs.Host}. The light-curves of SNe Ia in late-type host galaxies are significantly ($>$6.0$\sigma$) broader and slower declining than those in early-type hosts (see also Table~\ref{tab.Fitparameters.Vs.Host}). The faintest SNe Ia are found in early-type hosts, while the most brightest SNe are found in late-type hosts. When we look at the trend with more specific morphological types of hosts (the first panel in the figures), the difference in light-curve shape parameters between SNe in E-S0 and Scd/Sd/Irr hosts is $\sim55\%$ increased compared to when splitting the sample into early- and late-types. Considering the stellar population age difference between those morphological types, this result indirectly shows that there could be a different population age in the SN progenitors.

The light-curve shape parameters as a function of host $M_{stellar}$ and global $sSFR$ can be found in the third and fourth panels of the top panels of Figures~\ref{fig.S2.Fitparameters.Vs.Host} and \ref{fig.MLCS.Fitparameters.Vs.Host}. The figures show that broader and slower declining SNe Ia are more likely to be found in low-mass and globally star-forming hosts at a $>$6.5$\sigma$ confidence level (see also Table~\ref{tab.Fitparameters.Vs.Host}). Note that, in low-mass and high $sSFR$ hosts, narrower and faster declining SNe are rarer, while SNe in high-mass and low $sSFR$ have a much wider light-curve shape range. We also see evidence that the light-curve shape parameter is a continuous variable of $M_{stellar}$ and global $sSFR$, as pointed out by \citet{Sullivan2010}. 

\subsubsection{Supernova Color and Host Extinction: $C$ and $A_{V}$\label{sec.LCColorAv.Vs.Host}}

Lower panels of Figures~\ref{fig.S2.Fitparameters.Vs.Host} and \ref{fig.MLCS.Fitparameters.Vs.Host} show the trend of the SN Ia color for SALT2 and the host extinction for MLCS2k2 with host properties, respectively. The situation in interpreting SN Ia color is very complicated, as $C$ is a single parameter that captured both SN intrinsic color and host reddening by dust. Even though several studies discussed on this issue \citep[e.g.,][]{Hicken2009b, Neill2009, Sullivan2010, Childress2013, Pan2014}, there is no conclusive result yet. For the MLCS2k2 $A_{V}$, this is the first time in SN host studies to investigate the trend with host $M_{stellar}$ and $sSFR$.

The SN color trend with host properties can be found in the lower panels of Figure~\ref{fig.S2.Fitparameters.Vs.Host}. At first, we would expect that SNe Ia in late-type and globally star-forming host galaxies are redder, because these galaxies are thought to contain more dust than early-type and globally passive galaxies in general. However, no strong trends are found with host morphological types and global $sSFR$. On the other hand, we find a mild trend with the host stellar mass. SNe Ia in high-mass hosts show slightly bluer than those in low-mass hosts: the difference in the weighted mean of $C$ is 0.030 $\pm$ 0.008 ($3.8\sigma$, Table~\ref{tab.Fitparameters.Vs.Host}). We note that the reddest SNe Ia (e.g., $C\geq0.2$) prefer high-mass and globally star-forming hosts.

In the lower panels of Figure~\ref{fig.MLCS.Fitparameters.Vs.Host}, we plot the dust extinction value for host galaxies estimated from MLCS2k2 as a function of host properties. As expected, we find that late-type and globally star-forming host galaxies show $0.117\pm0.032$ mag ($3.7\sigma$) and $0.035\pm0.017$ mag ($2.1\sigma$) higher extinction value than early-type and globally passive hosts, respectively (see Table~\ref{tab.Fitparameters.Vs.Host}). The difference is the largest ($0.191\pm0.045$ mag, $4.2\sigma$) when we split the hosts into more specific morphological types. In contrary to SALT2 $C$, there is no trend with $M_{stellar}$.

\subsection{Luminosity Dependence of Type Ia Supernovae on the Properties of Host Galaxies\label{sec.3.3}}

\begin{table*}
\centering
\caption{\label{tab.HR.Vs.Host} The Weighted Mean and rms Scatter of Hubble Residuals in Different Host Properties}
\begin{adjustbox}{width=\textwidth,center=\textwidth}
\begin{tabular}{lccrccccccccrcccccc}
\toprule
Host     & Group & \multicolumn{8}{c}{SALT2} & & \multicolumn{8}{c}{MLCS2k2} \\
\cline{3-10} \cline{12-19}
Property &       & $N_{SN}$ & \multicolumn{1}{c}{HR$_{WM}$} && Error &&  rms  && Error & & $N_{SN}$ & \multicolumn{1}{c}{HR$_{WM}$} && Error &&  rms  && Error \\
         &       &          & \multicolumn{1}{c}{(mag)}     && (mag) && (mag) && (mag) & &          & \multicolumn{1}{c}{(mag)}     && (mag) && (mag) && (mag) \\
\midrule
Mass & High-mass & 464 & $-0.022$ && 0.008 && 0.172 && 0.006 && 366 & $-0.023$ && 0.009 && 0.166 && 0.006 \\
     & Low-mass  & 184 &   0.035  && 0.012 && 0.164 && 0.009 && 138 &   0.042 && 0.012 && 0.140 && 0.008 \\
\hline
Diff. & & & \textbf{0.057} && \textbf{0.014} &&  &&  && & \textbf{0.065} && \textbf{0.015} &&  &&  \\
\hline
\\
\hline
sSFR & Globally Passive      & 194 & $-0.043$ && 0.013 && 0.180 && 0.009 && 152 & $-0.034$ && 0.013 && 0.162 && 0.009 \\
     & Globally Star-Forming & 455 &   0.006  && 0.008 && 0.167 && 0.006 && 354 & $-0.001$ && 0.009 && 0.163 && 0.006 \\
\hline
Diff. & & & \textbf{0.049} && \textbf{0.015} &&  &&  && & \textbf{0.033} && \textbf{0.016} &&  &&  \\
\hline
\\
\hline
sSFR & Locally Passive      & 194 & $-0.043$ && 0.013 && 0.180 && 0.009 && 152 & $-0.034$ && 0.013 && 0.162 && 0.009 \\
     & Locally Star-Forming  & 174 &   0.038    && 0.013 && 0.172 && 0.009 && 129 &    0.038  && 0.012 && 0.138 && 0.009 \\
\hline
Diff. & & & \textbf{0.081} && \textbf{0.018} &&  &&  && & \textbf{0.072} && \textbf{0.018} &&  &&  \\
\hline
\\
\hline
Morphology & E-S0       &  44 & 0.036 && 0.036 && 0.220 && 0.022 &&  32 &   0.038  && 0.036 && 0.187 && 0.024 \\
           & S0a-Sc     & 131 & 0.001 && 0.018 && 0.191 && 0.012 && 103 & $-0.014$ && 0.019 && 0.176 && 0.012 \\
           & Scd/Sd/Irr &  16 & 0.018 && 0.038 && 0.119 && 0.013 &&  14 &   0.073  && 0.028 && 0.049 && 0.010 \\
\hline
Diff.      & (Scd/Sd/Irr - E-S0) & & \textbf{0.018} && \textbf{0.052} &&  &&  && & \textbf{0.035} && \textbf{0.046} &&  &&  \\
\hline
\\
\hline
Morphology & Early-type &  66 & $-0.012$ && 0.023 && 0.178 && 0.016 &&  50 & $-0.019$ && 0.024 && 0.164 && 0.017 \\
           & Late-type  & 177 & $-0.009$ && 0.014 && 0.173 && 0.009 && 143 &   0.001  && 0.015 && 0.162 && 0.010 \\
\hline
Diff.      & & & \textbf{0.003} && \textbf{0.027} &&  &&  && & \textbf{0.020} && \textbf{0.028} &&  &&   \\
\bottomrule
\end{tabular}
\end{adjustbox}
\end{table*}

\begin{figure}
\centering
\includegraphics[angle=0,width=\columnwidth]{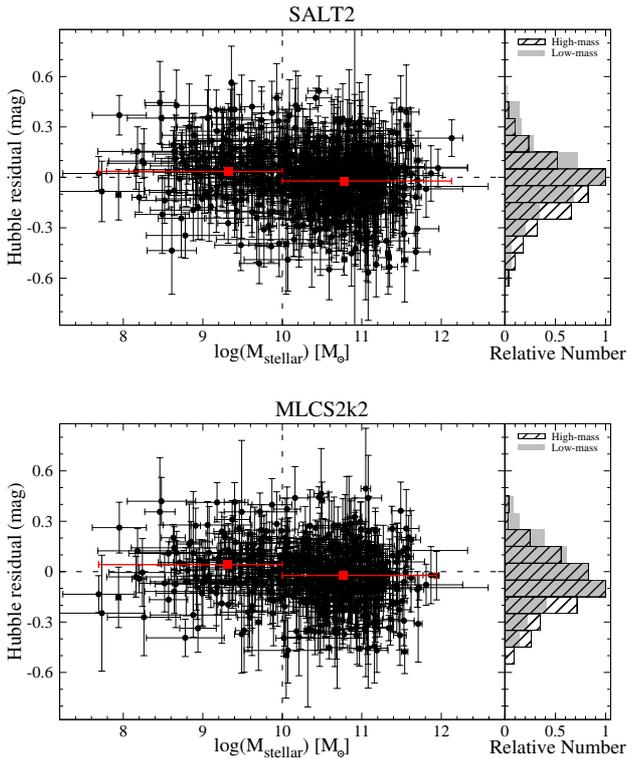}
\caption[Hubble residuals for SALT2 and MLCS2k2 versus stellar mass of host galaxies]{HRs for SALT2 (upper panel) and MLCS2k2 (lower panel) versus host $M_{stellar}$. We find that SNe Ia in the low-mass hosts are $\sim$0.06 mag fainter than those in the high-mass hosts. The red squares represent the weighted means of HRs in bins of host $M_{stellar}$. The vertical dotted line (log($M_{stellar}) = 10.0$) indicates the limit distinguishing between high-mass and low-mass host galaxies. The histogram on the right panel shows the HR distribution of SNe Ia in high-mass and low-mass hosts.}
\label{fig.HR.Vs.HostMass}
\end{figure}

\begin{figure}[t]
\centering
\includegraphics[angle=0,width=\columnwidth]{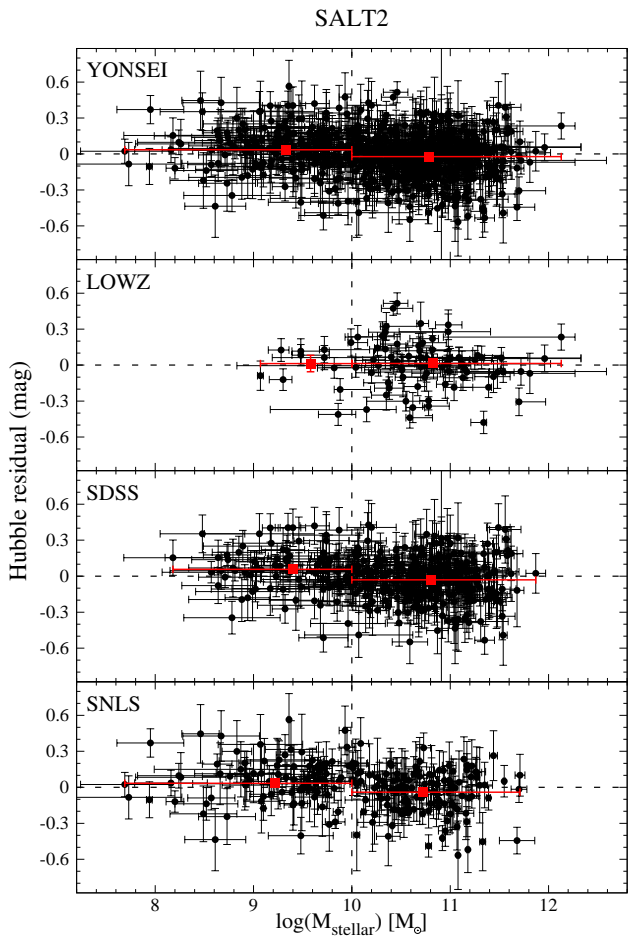}
\caption[SALT2 Hubble residuals versus stellar mass of host galaxies observed by surveys]{SALT2 HRs versus $M_{stellar}$ of host galaxies observed by surveys. Each subsample in the different redshift range follows the trend as we found with the YONSEI sample.}
\label{fig.S2HR.Vs.HostMass.bySurvey}
\end{figure}

\begin{figure}[t]
\centering
\includegraphics[angle=0,width=\columnwidth]{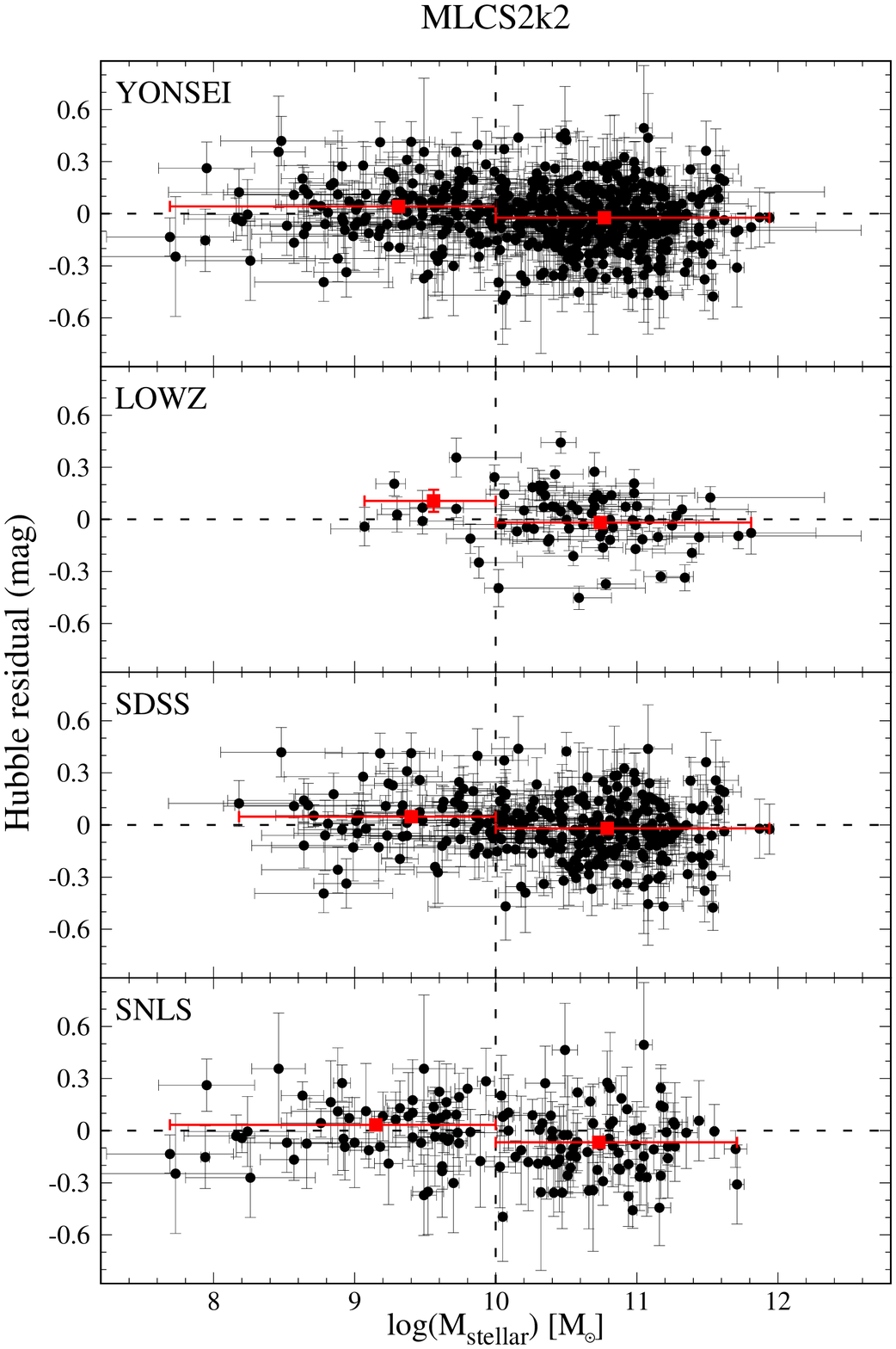}
\caption[MLCS2k2 Hubble residuals versus stellar mass of host galaxies by surveys]{Same as Figure~\ref{fig.S2HR.Vs.HostMass.bySurvey}, but for MLCS2k2 HRs.}
\label{fig.MLCSHR.Vs.HostMass.bySurvey}
\end{figure}

\begin{figure}[t]
\centering
\includegraphics[angle=0,width=\columnwidth]{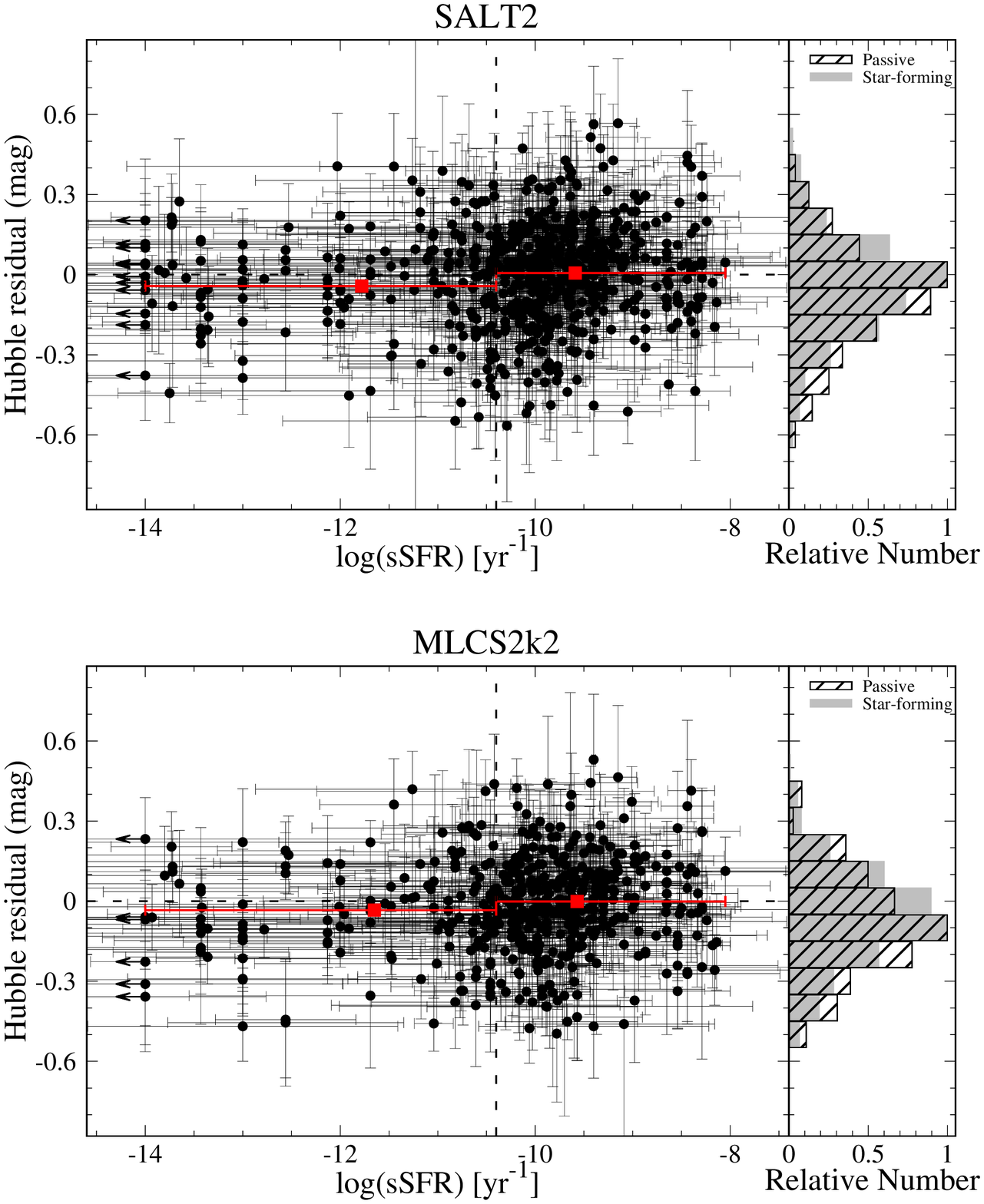}
\caption[Hubble residuals for SALT2 and MLCS2k2 versus global star formation rate of host galaxies]{Same as Figure~\ref{fig.HR.Vs.HostMass}, but for global $sSFR$ of host galaxies. We find that SNe Ia in the globally star-forming hosts are $\sim$0.04 mag fainter than those in the globally passive hosts. The vertical dotted line (log($sSFR) = -10.4$) indicates the limit distinguishing between passive and star-forming host galaxies for our sample. The histogram on the right panel shows the HR distribution of SNe Ia in passive and star-forming hosts.}
\label{fig.HR.Vs.HostsSFR}
\end{figure}

As we observed in the above section, the trends between light-curve fit parameters and host properties require a correction for the light-curve shape and color or extinction. If the correction perfectly works, we would not be able to observe the dependence of ``corrected'' luminosity of SNe Ia on the properties of host galaxies. In order to investigate this environmental effect, we plot the corrected luminosity of SNe Ia as a function of host global properties in Figures~\ref{fig.HR.Vs.HostMass} through \ref{fig.HR.Vs.HostMorphology}. The weighted mean and rms scatter of HRs with host properties for our sample are listed in Table~\ref{tab.HR.Vs.Host}.

For this investigation, as we have briefly introduced in Section~\ref{sec.2.4.1}, we use the definition of the Hubble residual (HR) for the corrected luminosity of SNe Ia. HR is defined as HR $\equiv \mu_\mathrm{SN}-\mu_\mathrm{model}(z)$, so that a negative residual means that the distance determined from the SN is less than the distance derived from the host galaxy redshift, together with the cosmological model. Therefore, in general, we interpret that the brighter SNe, after light-curve corrections, have negative HRs, and vice versa for the fainter SNe. In the calculation of the weighted mean of HRs in each host group, the error of the weighted mean is corrected to ensure a $\chi^2_{red} = 1$. We have also applied Chauvenet's criterion to reject outliers during this procedure. For our analysis, this criterion corresponds to $2.9\sigma$ for SALT2 and $2.8\sigma$ for MLCS2k2, on average.

\subsubsection{Host Stellar Mass\label{sec.Host.Mass}}

We start wih a well-establisehd correlation between SN HRs and host $M_{stellar}$. In Figure~\ref{fig.HR.Vs.HostMass}, the dependence of SN Ia luminosity on the host $M_{stellar}$ is shown for SALT2 (upper panel) and MLCS2k2 (lower panel). We find that SNe Ia in the low-mass hosts are fainter than those in the high-mass hosts: the difference in the weighted mean of HRs is $0.057\pm0.014$ mag ($4.1\sigma$) for SALT2 and $0.065\pm0.015$ mag ($4.3\sigma$) for MLCS2k2 (see Table~\ref{tab.HR.Vs.Host}). Our findings are quantitatively and qualitatively well consistent with the previous studies \citep[e.g.,][]{Kelly2010, Lampeitl2010, Sullivan2010}.

We also investigate the difference in HRs for LOWZ (low-redshift), SDSS (intermediate-redshift), and SNLS (high-redshift) samples separately. In Figures~\ref{fig.S2HR.Vs.HostMass.bySurvey} and \ref{fig.MLCSHR.Vs.HostMass.bySurvey}, the dependence of SN Ia luminosity on the host $M_{stellar}$ split by SN data is shown for SALT2 and MLCS2k2, respectively. Each subsample in the different redshift range follows the trend as we found with the YONSEI sample, and their HR differences are consistent within the $\sim$1$\sigma$ level (see Table~\ref{tab.HRs.Vs.Host.bySurvey}). This would indicate that the SN Ia luminosity dependence on the host $M_{stellar}$ is the global phenomenon over the whole redshift range. We note that because the LOWZ sample was discovered in host-targeted surveys which prefer more luminous and massive galaxies, as we pointed out in Section~\ref{sec.2.5}, there is a large imbalance between the fraction of low-mass and high-mass host galaxies (e.g., 15:85, compared to 31:69 for other samples on average). This may affect the trend in the LOWZ sample.

Finally, we compare the rms scatters for SNe Ia in the high-mass and those in the low-mass samples. SNe in low-mass hosts show $\sim$5$\%$ and $\sim$16$\%$ smaller rms scatter than those in high-mass hosts for SALT2 and MLCS2k2, respectively (see Table~\ref{tab.HR.Vs.Host}). This result appears to be that SNe Ia in the low-mass hosts are better standard candles than those in the high-mass hosts.

\subsubsection{Host Global Specific Star Formation Rate\label{sec.Host.sSFR}}

We plot the dependence of SN Ia luminosity on the host global $sSFR$ in Figure~\ref{fig.HR.Vs.HostsSFR}. We find that SNe Ia in the globally star-forming hosts are $0.049\pm0.015$ mag ($3.3\sigma$) and $0.033\pm0.016$ mag ($2.1\sigma$) fainter than those in the globally passive hosts for SALT2 and MLCS2k2, respectively (see Table~\ref{tab.HR.Vs.Host}). Our result is qualitatively consistent with previous results, but the HR differences are reduced \citep[e.g., $\sim0.100$ mag, see][]{Lampeitl2010, Sullivan2010, D'Andrea2011}. We note that the weighted mean of HR for SNe Ia in globally star-forming hosts interestingly shows $\sim$0.0 mag, which implies that those SNe can be a good anchor for the Hubble--Lem\^aitre diagram.

\begin{figure}[t]
\centering
\includegraphics[angle=0,width=\columnwidth]{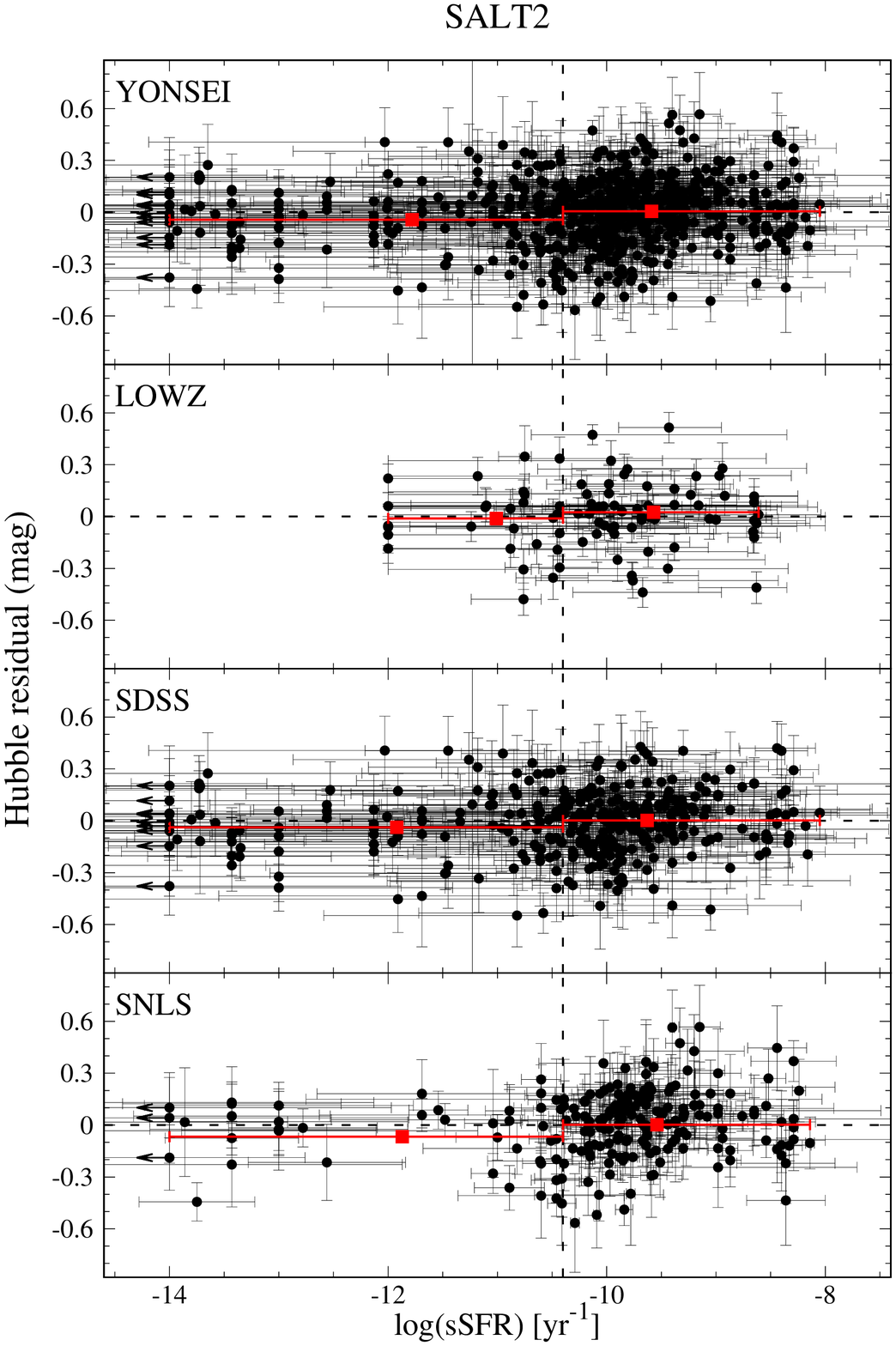}
\caption[SALT2 Hubble residuals versus global star formation rate of host galaxies observed by surveys]{SALT2 HRs versus global $sSFR$ of host galaxies observed by surveys. The trends in each subsample in the different redshift range agree well with the trend observed in the YONSEI sample.}
\label{fig.S2HR.Vs.HostsSFR.bySurvey}
\end{figure}

\begin{figure}[t]
\centering
\includegraphics[angle=0,width=\columnwidth]{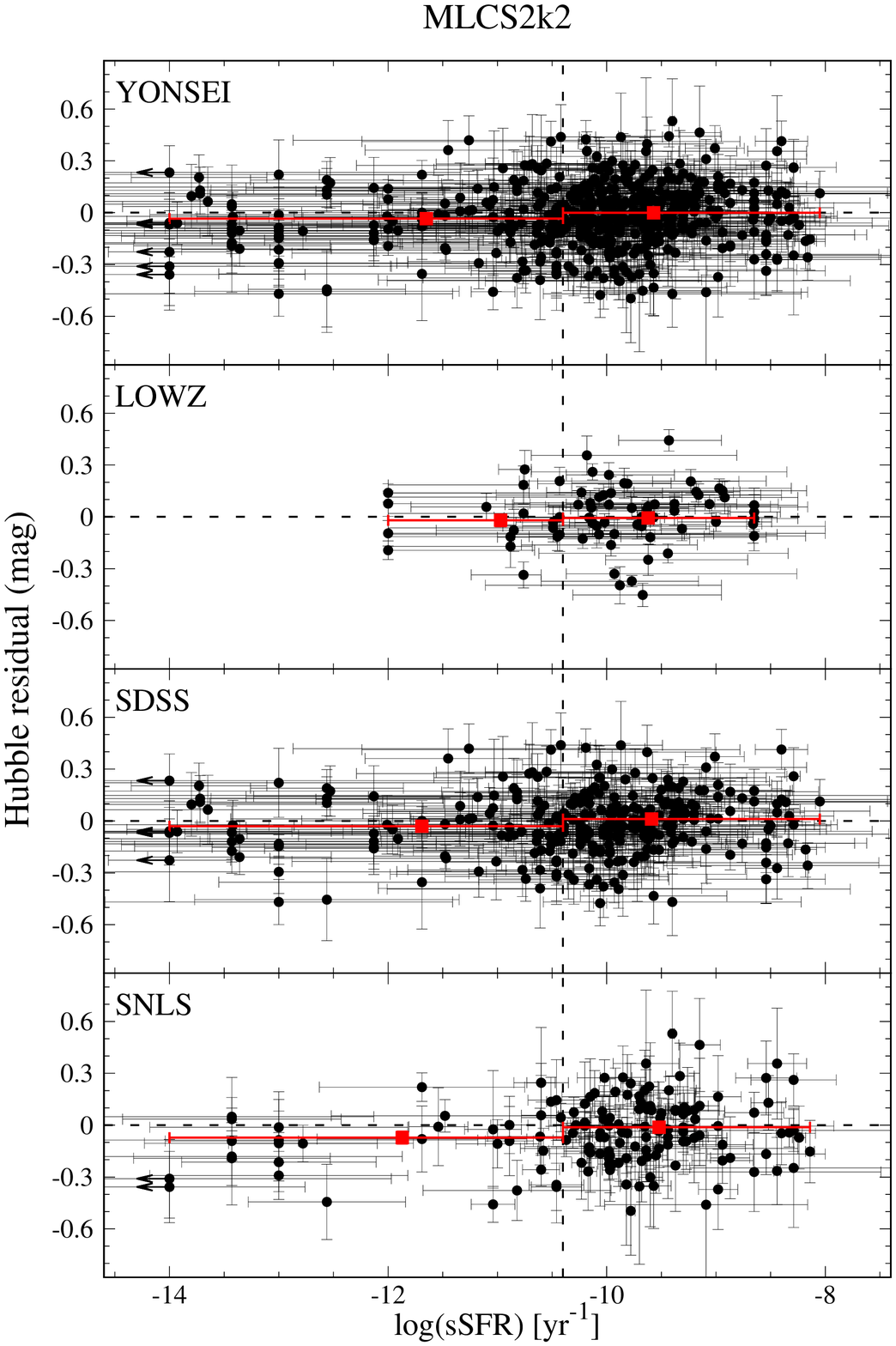}
\caption[MLCS2k2 Hubble residuals versus global star formation rate of host galaxies by surveys]{Same as Figure~\ref{fig.S2HR.Vs.HostsSFR.bySurvey}, but for MLCS2k2 HRs.}
\label{fig.MLCSHR.Vs.HostsSFR.bySurvey}
\end{figure}

In Figures~\ref{fig.S2HR.Vs.HostsSFR.bySurvey} and \ref{fig.MLCSHR.Vs.HostsSFR.bySurvey}, the SN Ia luminosity trends with host global $sSFR$ split by SN data are shown for SALT2 and MLCS2k2, respectively. The trends in each subsample in the different redshift range agree well with the trend observed in the YONSEI sample, even though at slightly lower significance (see Table~\ref{tab.HRs.Vs.Host.bySurvey}). As the mean $sSFR$ is known to increase with redshift \citep[e.g.,][]{Madau2014, Driver2018}, the fraction of globally star-forming hosts in our sample is also increased by $\sim$10$\%$ with redshift. In addition, the overall SN Ia occurrence rate of globally star-forming hosts in the YONSEI sample is $\sim$2.3 times larger than that of globally passive hosts, and this is qualitatively consistent with the results of \citet{Sullivan2006} and \citet{Smith2012}.

The rms scatter of SNe Ia in globally star-forming hosts shows a $\sim$7$\%$ less scatter than those in globally passive hosts for SALT2. However, we see no difference in the rms scatter for the MLCS2k2 global $sSFR$ sample. Previous studies also reached different conclusions for each author. For example, \citet{Kelly2015}, \citet{Uddin2017}, and \citet{Kim2018} supported that SNe Ia in the star-forming sample show a less scatter, while \citet{Lampeitl2010}, \citet{Sullivan2010}, and \citet{Jones2015} found the opposite result.

\begin{figure}[t]
\centering
\includegraphics[angle=0,width=0.9\columnwidth]{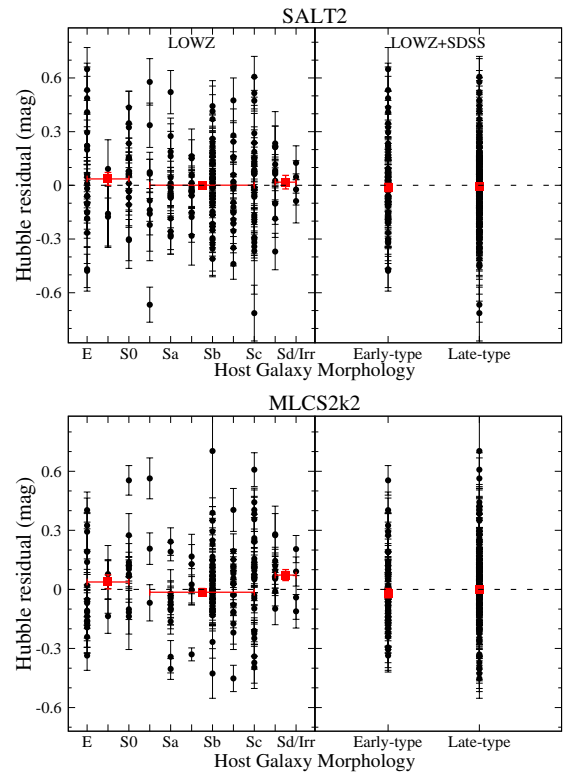}
\caption[Hubble residuals for SALT2 and MLCS2k2 vs. host galaxy morphology]{Same as Figure~\ref{fig.HR.Vs.HostMass}, but for the host galaxy morphology. We observe no conclusive trend.}
\label{fig.HR.Vs.HostMorphology}
\end{figure}

\begin{table*}
\centering
\caption{\label{tab.HRs.Vs.Host.bySurvey} Summary of Hubble Residual Differences in Each Data Set}
\begin{adjustbox}{width=0.8\textwidth,center=\textwidth}
\begin{tabular}{lccccccc}
\hline\hline
        & \multicolumn{7}{c}{SALT2} \\
\cline{2-8}
SN Data & \multicolumn{3}{c}{Mass} && \multicolumn{3}{c}{sSFR} \\
\cline{2-4} \cline{6-8}
        & $N_{SN}$ & $N_{Low}:N_{High}$ & HR Diff.       && $N_{SN}$ & $N_{Pas}:N_{SF}$ & HR Diff.  \\
        &          &                    & (mag)          &&          &                  & (mag)     \\
\hline
\textbf{YONSEI} & 648 & 184:464 & $0.057\pm0.014$ ($4.1\sigma$) && 649 & 194:455 & $0.049\pm0.015$ ($3.3\sigma$) \\
\hline
LOWZ            & 88  & 12:76   & $0.002\pm0.072$ ($0.0\sigma$) && 86  & 28:59   & $0.054\pm0.044$ ($1.2\sigma$) \\
SDSS            & 350 & 87:263  & $0.088\pm0.018$ ($4.9\sigma$) && 350 & 122:228 & $0.040\pm0.019$ ($2.1\sigma$) \\
SNLS            & 207 &85:122   & $0.074\pm0.023$ ($3.2\sigma$) && 207 & 45:162  & $0.069\pm0.031$ ($2.2\sigma$) \\
\hline
\\
\hline \hline
      & \multicolumn{7}{c}{MLCS2k2} \\
\cline{2-8}
SN Data & \multicolumn{3}{c}{Mass} && \multicolumn{3}{c}{sSFR} \\
\cline{2-4} \cline{6-8}
        & $N_{SN}$ & $N_{Low}:N_{High}$ & HR Diff.       && $N_{SN}$ & $N_{Pas}:N_{SF}$ & HR Diff.  \\
        &          &                    & (mag)          &&          &                  & (mag)     \\
\hline
\textbf{YONSEI} & 504 & 138:366 & $0.065\pm0.015$ ($4.3\sigma$) && 506 & 152:354 & $0.033\pm0.016$ ($2.1\sigma$) \\
\hline
LOWZ            & 71  & 11:60   & $0.124\pm0.068$ ($1.8\sigma$) && 72  & 18:54   & $0.013\pm0.042$ ($0.3\sigma$) \\
SDSS            & 288 & 69:219  & $0.066\pm0.019$ ($3.5\sigma$) && 287 & 99:188  & $0.042\pm0.019$ ($2.2\sigma$) \\
SNLS            & 140 & 57:83   & $0.102\pm0.025$ ($4.1\sigma$) && 141 & 36:105  & $0.061\pm0.037$ ($1.6\sigma$) \\
\hline
\end{tabular}
\end{adjustbox}
\end{table*}

\subsubsection{Host Morphology\label{sec.Host.Morphology}}

Figure~\ref{fig.HR.Vs.HostMorphology} shows the dependence of SN Ia luminosity on the host galaxy morphology for SALT2 (upper panel) and MLCS2k2 (lower panel). We observe no conclusive trends for our LOWZ and intermediate-redshift SDSS samples. More quantitatively, the HR difference is $0.011\pm0.029$ mag and $0.028\pm0.027$ mag on average for SALT2 and MLCS2k2, respectively (see Table~\ref{tab.HR.Vs.Host}). The results show a greatly reduced difference in HRs compared to the previous results, which found the HR difference of $0.144\pm0.070$ mag for the low-redshift sample \citep[$z < 0.1$;][]{Hicken2009b} and $0.180\pm0.090$ mag for the high-redshift sample \citep[$z > 0.9$;][]{Suzuki2012}.

We also find that SNe Ia in Scd/Sd/Irr hosts have the lowest rms scatter ($\sim$0.084 mag, see Table~\ref{tab.HR.Vs.Host}), which is consistent with the result of \citet{Hicken2009b}. These SNe show a $46\%$ and $74\%$ smaller rms scatter than those in E-S0 hosts for SALT2 and MLCS2k2, respectively. This may imply that SNe Ia exploded in Scd/Sd/Irr hosts would provide more robust results in estimating cosmological parameters.

\subsubsection{Comparison with Previous Studies\label{sec.Comparison}}

\begin{figure}
\centering
\includegraphics[angle=0,width=0.9\columnwidth]{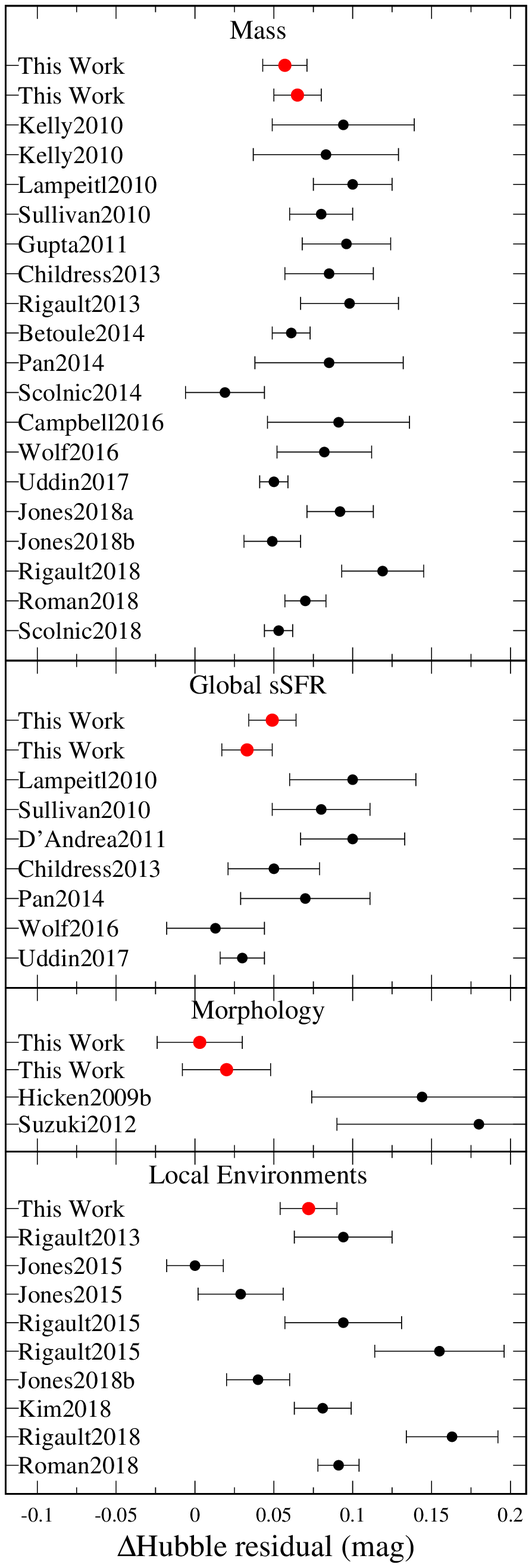}
\caption[Comparison of Hubble Residual Differences between Previous Studies]{Comparison of HR differences between previous studies. We find that our results are consistent with the results reported by many previous studies. See Table~\ref{tab.Comparison} for more information about each study.}
\label{fig.HR.Comparison}
\end{figure}

In this section, we summarize our findings of the difference in SN Ia HRs in different host galaxy properties and compare them with the results reported in previous studies. We present a summary of our findings and previous results in Table~\ref{tab.Comparison} and Figure~\ref{fig.HR.Comparison}. We note that we compare our results only with those which used corrections for both the light-curve shape and SN color or host extinction (e.g, after 2007, when \citealt{Guy2007} and \citealt{Jha2007} released SALT2 and MLCS2k2 fitters, respectively).

We first compare our result with previous studies with respect to host $M_{stellar}$. As shown in the table, there are many studies with a variety of samples over a large redshift range. They found that SNe Ia in high-mass hosts appear brighter than those in low-mass hosts, after light-curve corrections with several light-curve fitters. The size of HR difference is 0.08 mag on average and the transition mass is around $10^{10}$ M$_{\odot}$. Our result indeed is fully consistent with previous studies within the $\sim$1$\sigma$ level.

Several points in the table for the mass comparison are worth noting here. The first is that some of the recent studies tried to include photometrically classified SNe Ia in their sample \citep[e.g.,][]{Campbell2016, Uddin2017, Jones2018a}. With this inclusion, their sample size is $\sim$2 times larger than other studies. The error of the HRs, however, is not small as much as expected from the sample size, because of the contamination from other types of SNe. The second one of note regards the light-curve fitting method for the results of \citet{Jones2018a} and \citet{Scolnic2018}. They employed the same equation for the SALT2 SN distance modulus as we used (see Equation~\ref{eq.DM_SN}), but they included two extra terms: $\Delta_{M}$ for the host mass dependence correction and $\Delta_{B}$ for the simulated selection bias correction (see Equation~\ref{eq.DM_SN_S17}). Interestingly, even though they considered the HR difference among different host masses prior to estimating the distance modulus of SNe, the dependence of SN Ia luminosity on the host mass is still observed at $\sim$5.2$\sigma$ level. The last note is that \citet{Scolnic2014} found no trend between HRs and host stellar mass. However, the interesting point is that with a $\sim$10 times larger sample which includes \citet{Scolnic2014} data, the trend is recovered in \citet{Scolnic2018}.

We now turn to the results of host global $sSFR$. Our result is consistent with previous studies in the sense that SNe Ia in passive hosts are brighter than those in star-forming hosts. The size of HR difference is 0.06 mag on average. However, the size of difference is smaller and shows statistically less significant than the result split by host $M_{stellar}$.

Assuming that host $M_{stellar}$ and $sSFR$ are closely related to the galaxy morphology, we expect from above that brighter SNe Ia would be exploded in the early-type host galaxies. However, we find no HR difference between SNe Ia in early-type hosts and those in late-type hosts, while previous studies found at $\sim$2$\sigma$ level. This may be because our LOWZ sample, which is dominant in the host morphology study ($\sim$80$\%$), is drawn from a host-targeted survey, as pointed out in Section~\ref{sec.2.5}. It leads to that most of the hosts in the LOWZ sample are in the high-mass region (see the second panels of Figures~\ref{fig.S2HR.Vs.HostMass.bySurvey} and \ref{fig.MLCSHR.Vs.HostMass.bySurvey}), thus there would be no HR difference between SNe in early-type hosts and those in late-type hosts. 

Finally, we look at the trends with local environments of SNe Ia. Our finding which includes a sample at high-redshift range shows a good agreement with previous results. They concluded that SNe Ia occurring locally passive and redder in local $U-V$ color are brighter than those in locally star-forming and bluer color environments, except the results of \citet{Jones2015}\footnote{This is because of the effect of the redshift cut they applied for their sample. \citet{Kim2018} recovered the \citet{Jones2015} and \citet{Rigault2015} results, which show an apparent discrepancy, by applying the same redshift cut used by each of them.}. The size of HR difference is $\sim$0.1 mag, which is a larger difference than when using host global properties. As the local environment is more directly linked to the SN progenitor, the results strongly suggest that there are different populations in SN Ia progenitor.

From this comparison, we conclude that our results are consistent with the results reported by many previous studies. However, our findings are an independent confirmation based on the spectroscopically confirmed SNe Ia from a combined sample of LOWZ, SDSS, and SNLS data, together with two different light-curve fitters.

\begin{table*}
\centering
\caption{\label{tab.Comparison} Comparison of Hubble Residual Differences between Previous Studies}
\begin{adjustbox}{width=1.1\textwidth,center=\textwidth}
\begin{tabular}{lccccc}
\toprule
 Study & SN Data & $N_{SN}$ & Redshift & HR Diff.  & LC Fitter \\
       &         &          & Range    & (mag)    &          \\
\midrule
\\
Mass &&&&& \\
\hline
\textbf{This Work}             & YONSEI            & 648        & $0.01 < z < 0.85$  & $0.057\pm0.014$ ($4.1\sigma$) & SALT2       \\
\textbf{This Work}             & YONSEI            & 504        & $0.01 < z < 0.85$  & $0.065\pm0.015$ ($4.3\sigma$) & MLCS2k2 ($R_V = 2.2$)     \\
\citet{Kelly2010}     & CfA               &  62        & $0.015 < z < 0.08$ & $0.094\pm0.045$ ($2.1\sigma$) & SALT2       \\
\citet{Kelly2010}     & CfA               &  60        & $0.015 < z < 0.08$ & $0.083\pm0.046$ ($1.8\sigma$) & MLCS2k2     \\
\citet{Lampeitl2010}  & SDSS              & 162        & $0.05 < z < 0.21$  & $0.100\pm0.025$ ($4.0\sigma$) & SALT2       \\
\citet{Sullivan2010}  & SNLS              & 195        & $0.01 < z < 0.85$  & $0.080\pm0.020$ ($4.0\sigma$) & SALT2+SiFTO \\
\citet{Gupta2011}     & SDSS              & 206        & $0.01 < z < 0.42$  & $0.096\pm0.028$ ($3.4\sigma$) & SALT2       \\
\citet{Childress2013} & SNf               & 115        & $0.03 < z < 0.08$  & $0.085\pm0.028$ ($3.0\sigma$) & SALT2       \\
\citet{Rigault2013}   & SNf               &  82        & $0.03 < z < 0.08$  & $0.098\pm0.031$ ($3.2\sigma$) & SALT2       \\	
\citet{Betoule2014}   & JLA               & 740        & $0.01 < z < 1.4$   & $0.061\pm0.012$ ($5.1\sigma$) & SALT2       \\
\citet{Pan2014}       & PTF               &  50        & $0.01 < z < 0.09$  & $0.085\pm0.047$ ($1.8\sigma$) & SiFTO       \\
\citet{Scolnic2014}   & PS                & 110        & $0.03 < z < 0.65$  & $0.019\pm0.025$ ($0.8\sigma$) & SALT2       \\
\citet{Campbell2016}  & SDSS              & $581^{a}$  & $0.05 < z < 0.55$  & $0.091\pm0.045$ ($2.0\sigma$) & SALT2       \\
\citet{Wolf2016}      & SDSS              & 144        & $0.05 < z < 0.3$   & $0.082\pm0.030$ ($2.7\sigma$) & SALT2       \\
\citet{Uddin2017}     & CfA+CSP+SDSS+SNLS & $1338^{b}$ & $0.01 < z < 1.1$   & $0.050\pm0.009$ ($5.6\sigma$) & SALT2                         \\
\citet{Jones2018a}     & CfA+CSP+PS        & $1369^{c}$ & $0.01 < z < 0.7$   & $0.092\pm0.021$ ($4.4\sigma$) & SALT2 with $\Delta_{M}$ and $\Delta_{B}$$^{d}$ \\
\citet{Jones2018b}     & Pantheon+Foundation        & 216 & $0.01 < z < 0.1$   & $0.049\pm0.018$ ($2.7\sigma$) & SALT2 with $\Delta_{B}$$^{d}$ \\
\citet{Rigault2018}            & SNf  & 141  & $0.02 < z < 0.08$  & $0.119\pm0.026$ ($4.6\sigma$) & SALT2      \\
\citet{Roman2018}     & CfA+CSP+SDSS+SNLS & 666        & $0.01 < z < 0.8$   & $0.070\pm0.013$ ($5.4\sigma$) & SALT2                         \\
\citet{Scolnic2018}   & Pantheon          & 1023       & $0.01 < z < 2.3$   & $0.053\pm0.009$ ($5.9\sigma$) & SALT2 with $\Delta_{M}$ and $\Delta_{B}$$^{d}$ \\
\hline
&&&&& \\
Global Specific Star Formation Rate &&&&& \\
\hline
\textbf{This Work}             & YONSEI            & 649        & $0.01 < z < 0.85$  & $0.049\pm0.015$ ($3.3\sigma$) & SALT2   \\
\textbf{This Work}             & YONSEI            & 506        & $0.01 < z < 0.85$  & $0.033\pm0.016$ ($2.1\sigma$) & MLCS2k2 ($R_V = 2.2$) \\
\citet{Lampeitl2010}  & SDSS              & 162        & $0.05 < z < 0.21$  & $0.100\pm0.040$ ($2.5\sigma$) & SALT2       \\
\citet{Sullivan2010}  & SNLS              & 195        & $0.01 < z < 0.85$  & $0.080\pm0.031$ ($2.6\sigma$) & SALT2+SiFTO \\
\citet{D'Andrea2011}  & SDSS              &  55        & $z < 0.15$         & $0.100\pm0.033$ ($3.0\sigma$) & SALT2       \\
\citet{Childress2013} & SNf               & 115        & $0.03 < z < 0.08$  & $0.050\pm0.029$ ($1.7\sigma$) & SALT2       \\
\citet{Pan2014}       & PTF               &  48        & $0.01 < z < 0.09$  & $0.070\pm0.041$ ($1.7\sigma$) & SiFTO       \\
\citet{Wolf2016}      & SDSS              & 144        & $0.05 < z < 0.3$   & $0.013\pm0.031$ ($0.5\sigma$) & SALT2       \\
\citet{Uddin2017}     & CfA+CSP+SDSS+SNLS & $1338^{c}$ & $0.01 < z < 1.1$   & $0.030\pm0.014$ ($2.1\sigma$) & SALT2       \\
\hline
\\
Morphology &&&&& \\
\hline
\textbf{This Work}           & YONSEI   & 243 & $0.01 < z < 0.2$  & $0.003\pm0.027$ ($0.1\sigma$) & SALT2         \\
\textbf{This Work}           & YONSEI   & 193 & $0.01 < z < 0.2$  & $0.020\pm0.028$ ($0.7\sigma$) & MLCS2k2 ($R_V = 2.2$)       \\
\citet{Hicken2009b} & CfA      &  97 & $0.01 < z < 0.1$  & $0.144\pm0.070$ ($2.1\sigma$) & SALT2+MLCS2k2 \\
\citet{Suzuki2012}  & Union2.1 &  28 & $0.9 < z < 1.5$   & $0.180\pm0.090$ ($2.0\sigma$) & SALT2         \\
\hline
\\
Local Environments \\
\hline
\textbf{This Work}            & YONSEI  & 281  & $0.01 < z < 0.85$  & $0.072\pm0.018$ ($4.0\sigma$) & MLCS2k2 ($R_V = 2.2$)  \\
\citet{Rigault2013}            & SNf         & 82   & $0.03 < z < 0.08$  & $0.094\pm0.031$ ($4.5\sigma$) & SALT2      \\
\citet{Jones2015}             & CfA+CSP+CT+SDSS+SNLS+PS1  & 179  & $0.01 < z < 0.1$  & $0.000\pm0.018$ ($0.0\sigma$) & SALT2      \\
\citet{Jones2015}             & CfA+CSP+CT+SDSS+SNLS+PS1  & 156  & $0.01 < z < 0.1$  & $0.029\pm0.027$ ($1.1\sigma$) & MLCS2k2 ($R_V = 2.5$)     \\
\citet{Rigault2015}            & CfA  & 77  & $0.023 < z < 0.1$  & $0.094\pm0.037$ ($2.5\sigma$) & SALT2      \\
\citet{Rigault2015}            & CfA  & 81  & $0.023 < z < 0.1$  & $0.155\pm0.041$ ($3.8\sigma$) & MLCS2k2 ($R_V = 2.5$)      \\
\citet{Jones2018b}           & Pantheon+Foundation   & 195  & $0.01 < z < 0.1$  & $0.040\pm0.020$ ($2.0\sigma$) & SALT2 with $\Delta_{B}$$^{d}$      \\
\citet{Kim2018}                & YONSEI  & 368  & $0.01 < z < 0.85$  & $0.081\pm0.018$ ($4.5\sigma$) & SALT2      \\
\citet{Rigault2018}            & SNf  & 141  & $0.02 < z < 0.08$  & $0.163\pm0.029$ ($5.6\sigma$) & SALT2      \\
\citet{Roman2018}           & CfA+CSP+SDSS+SNLS & 666        & $0.01 < z < 0.8$   & $0.091\pm0.013$ ($7.0\sigma$) & SALT2                         \\
\bottomrule
\end{tabular}
\end{adjustbox}
\tabnote{
a. 581 photometrically classified SNe Ia.
\\b. 755 photometrically classified SNe Ia are included in their SDSS and SNLS samples.
\\c. 1035 photometrically classified SNe Ia are included in their PS sample.
\\d. $\Delta_{M}$ is a distance correction based on the host mass and $\Delta_{B}$ is another distance correction based on the predicted selection bias estimated from SN survey simulations.
}
\end{table*}

\section{Estimating Cosmological Parameters from SNe Ia in Different Environments\label{sec.Estimating.Cosmology}}

We observed in the previous section that the luminosity of SNe Ia varies with their host properties. This would imply that using a whole sample of SNe Ia together without considering host environments can cause a bias in estimating cosmological parameters; for example, a $\sim$10$\%$ shift in $w$ and a 3.3$\%$ correction for $H_{0}$ \citep{Lampeitl2010, Sullivan2010, Sullivan2011, Rigault2013, Rigault2015, Campbell2016, Uddin2017}. In order to investigate this bias in the YONSEI Cosmology sample, here we split the SN Ia sample according to the host environments as following the previous section (e.g., see Table~\ref{tab.HR.Vs.Host}), and then estimate cosmological parameters separately.

\begin{table*}
\centering
\caption{\label{tab:cosmology} Best-Fit Flat $\Lambda$CDM Parameters Estimated from SNe Ia in Different Host Environments}
\begin{tabular}{lcccccccc}
\toprule
Group && SNe & $\Omega_{M}$ & $\alpha$ & $\beta$ & $M_{B}$ & $\sigma_{int}$ & $\chi^{2}$/D.O.F. \\
\midrule
High-Mass             && 464 & $0.32^{+0.12}_{-0.08}$ & $0.15^{+0.01}_{-0.02}$ & $3.26\pm0.20$          & $-19.07^{+0.03}_{-0.02}$ & 0.112 & $458.23/460$ \\
Low-Mass              && 184 & $0.29^{+0.09}_{-0.06}$ & $0.15\pm0.03$          & $3.28^{+0.36}_{-0.34}$ & $-19.03^{+0.05}_{-0.04}$ & 0.098 & $179.37/180$ \\
\hline
Globally Passive      && 194 & $0.33^{+0.09}_{-0.10}$ & $0.18^{+0.03}_{-0.02}$ & $2.96^{+0.31}_{-0.29}$ & $-19.12^{+0.04}_{-0.05}$ & 0.104 & $190.62/190$ \\
Globally Star-Forming && 455 & $0.30\pm0.05$          & $0.13^{+0.02}_{-0.01}$ & $3.28^{+0.19}_{-0.20}$ & $-19.06^{+0.02}_{-0.03}$ & 0.105 & $450.08/451$ \\
\hline
Locally Passive      && 194 & $0.33^{+0.09}_{-0.10}$ & $0.18^{+0.03}_{-0.02}$ & $2.96^{+0.31}_{-0.29}$ & $-19.12^{+0.04}_{-0.05}$ & 0.104 & $190.62/190$ \\
Locally Star-Forming && 174 & $0.31\pm0.09$          & $0.14^{+0.04}_{-0.03}$ & $3.40^{+0.40}_{-0.37}$ & $-19.02\pm0.05$          & 0.111 & $170.58/170$ \\
\hline
YONSEI Cosmology        && 1049 & $0.32^{+0.03}_{-0.04}$ & $0.14^{+0.01}_{-0.00}$ & $3.07^{+0.14}_{-0.15}$ & $-19.06^{+0.01}_{-0.02}$ & 0.131 & $1038.15/1045$ \\
\bottomrule
\end{tabular}
\end{table*}

We use the same JLA likelihood code as we described in Section~\ref{sec.2.4.1} to estimate the best-fit cosmological parameters from SALT2 SNe Ia alone. For our baseline cosmology, we assume the flat $\Lambda$CDM model. One difference from the Section~\ref{sec.2.4.1} is that here we simultaneously calculate $\sigma_{int}$, instead of setting $\sigma_{int}=0$. As $\sigma_{int}$ is the uncertainty that makes a $\chi^2_\mathrm{red}$ equals unity, this uncertainty is required when we determine the `best-fit' cosmological parameters. The best-fit parameters estimated from the YONSEI Cosmology sample in different environments are listed in Table~\ref{tab:cosmology}.

The table shows that shifts in $\Omega_{M}$ and $\alpha$ are negligible, as pointed out by \citet{Sullivan2011} and \citet{Uddin2017}. However, in $\beta$, we can see the significant shift, except when splitting the sample based on host $M_{stellar}$. The value of $\beta$ in passive environments is lower than that in star-forming environments. This trend is also observed in \citet{Lampeitl2010} with the SDSS sample and \citet{Sullivan2010} with the SNLS sample. 

In terms of $\sigma_{int}$ (Table~\ref{tab:cosmology}) and rms scatter of HRs (Table~\ref{tab.HR.Vs.Host}), SNe Ia in low-mass and star-forming environments provide more robust results when estimating cosmological parameters. For example, SNe in locally star-forming environments have a $\sim$18$\%$ smaller rms scatter, and also require a $\sim$15$\%$ smaller intrinsic scatter than those in the full YONSEI Cosmology sample. From this, we can consider that SNe Ia in low-mass and star-forming environments would have similar progenitor ages that represent the most homogeneous sample \citep[see also][]{Childress2014, Kelly2015, Kim2018}. 

From the results described in this section, we conclude that cosmological parameters seem to be biased when we do not consider the SN Ia environments. As more SNe Ia surveys are coming with a much larger sample than ever, it would be expected that we start to see an obvious difference in cosmological parameters estimated from SNe Ia \citep{Uddin2017}. This means that we may need a host-related-correction \citep[see e.g.,][]{Sullivan2011, Suzuki2012, Scolnic2018}. However, investigating the origin of the environmental dependence of SN Ia luminosity should be preceded, which leads us to a more accurate cosmology.

\section{Discussion\label{sec:discussion}}

The purpose of this paper is to investigate the dependence of SNe Ia luminosity on global and local properties of host galaxies, explore the origin of the dependence, and predict the impact of the luminosity dependence on the SN cosmology. For this, we have constructed an independent SN Ia catalog, which has 1231 spectroscopically confirmed SNe Ia and 674 host galaxy data over the redshift range of $0.01<z<1.37$ and includes two independent light-curve fitters of SALT2 and MLCS2k2. From this catalog, we find that SNe Ia in low-mass and star-forming environments are $0.062\pm0.009$ mag and $0.057\pm0.010$ mag fainter than those in high-mass and passive environments, after ``empirical'' light-curve corrections with SALT2 and MLCS2k2, respectively (see Table~\ref{tab.HR.Vs.Host} for our main results). When only local environments of SNe Ia are considered, the luminosity difference increases to $0.081\pm0.018$ mag for SALT2 and $0.072\pm0.018$ mag for MLCS2k2. Our finding is consistent with previous studies (see Table~\ref{tab.Comparison}). However, our result is an independent confirmation based on a combined sample of SNe Ia from LOWZ, SDSS, and SNLS surveys ($0.1<z\le0.85$), by using two different light-curve fitters of SALT2 and MLCS2k2.

\subsection{On the Origin of the Environmental Dependence of SN Ia Luminosity\label{sec.conclusion.origin}}

As shown in Section~\ref{sec.Estimating.Cosmology}, the luminosity dependence would impact on estimating cosmological parameters. It is therefore important to investigate the origin of the dependence in order to use an SN Ia as a more accurate standard candle. 

These remaining trends, after ``empirical'' light-curve corrections, would indicate that 1) there are intrinsic ``physical'' processes that we do not understand yet, or 2) we simply require a ``third parameter'' when we analyze SN Ia light-curves. The latter case was intensively performed by \citet{Scolnic2018}. 

\citet{Scolnic2018} analyzed the SN light-curves using a modified version of SALT2. As we briefly described in Section~\ref{sec.Comparison}, they included two extra terms in their SALT2 analysis: $\Delta_{M}$ and $\Delta_{B}$, such that 

\begin{equation}
\mu_{SN} = m_{B} - M_{B} + \alpha \times X_{1} - \beta \times C + \Delta_{M} + \Delta_{B},
\label{eq.DM_SN_S17}
\end{equation}
where $\Delta_{M}$ is a distance correction based on the observed trend between SN Ia luminosity and host mass, and $\Delta_{B}$ is another distance correction based on the predicted bias estimated from SN survey simulations \citep{Kessler2017}. Furthermore, they considered distance biases due to intrinsic scatter, introduced by \citet{Scolnic2016}. However, after these all extra corrections, they still observed the HR difference of $0.053\pm0.009$ mag ($5.9\sigma$, see Table~\ref{tab.Comparison}). They argued that when they do not include these additional corrections the size of HR difference increases to $0.071\pm0.010$ mag ($7.1\sigma$). This result would indicate that the correction with various functional forms is not appropriate to capture the remaining trends in SN Ia luminosity and host galaxy properties. Instead of that, we require the understanding of detailed physics of SNe Ia.

Since the host $M_{stellar}$ and $sSFR$ cannot directly affect the SN luminosity, the theoretical studies suggested that the leading candidates for the observed trends are the progenitor age and the progenitor metallicity \citep{Timmes2003, Kasen2009}. Both are empirically known to correlate with the host $M_{stellar}$ and $sSFR$ \citep[e.g.,][]{Tremonti2004, Gallazzi2005, Kang2016}. 

In order to explore these issues more directly, there are several studies to deal with the host galaxy age and (gas-phase or stellar) metallicity \citep[e.g.,][]{Neill2009, Sullivan2010, Gupta2011, Johansson2013, Pan2014, Campbell2016, Wolf2016}. Most of the studies used the SED fitting technique or emission lines. However, the limitations of those methods, such as the age-metallicity degeneracy and attenuation by dust, are well-known \citep[see][]{Worthey1994, Walcher2011}. In order to overcome these limitations, Balmer absorption lines have been widely used in estimating the age and the metallicity of early-type galaxies during the last two decades \citep{Faber1992, Worthey1994, Worthey1998, Trager2000, Thomas2005, Kuntschner2006, Graves2007, Graves2009, Conroy2010}. In the recent study of \citet{Kang2016}, they employed Balmer absorption lines to determine more reliable population ages and metallicities for 27 early-type host galaxies. From high signal-to-noise observed spectra ($\geq$100 per pixel), they suggested that the stellar population age is mainly responsible for the relation between SN Ia luminosities and host properties at the $\sim$3.9$\sigma$ level. Even though more data are required to further confirm this result, this kind of study would provide the direct evidence for the origin of the environmental dependence and the evolution of SN Ia luminosity.

Furthermore, we would suggest another indirect approach to explore the origin of the environmental dependence of SN Ia luminosities in the context of a galaxy. Our results and many previous studies showed that SN Ia luminosity is changed at log($M_{stellar}$) = 10 (Section~\ref{sec.3.3}), and also as mentioned in Section~\ref{sec.2.5.1}, \citet{Kim2018} method to infer the local environments from the global host properties requires this mass scale. The mass scale of $10^{10} M_{\odot}$ takes on a unique position in galaxy studies. Numerous observational studies found that a transition in the assembly histories of galaxies for both early- and late-types and a transition of galaxy morphology occur near this mass scale \citep[e.g.,][]{Kauffmann2003, Balcells2007, Hopkins2009, Cappellari2013, Bernardi2014}. In addition, several recent simulations predicted another transition between the SN feedback and the AGN feedback at near this mass \citep[e.g.,][]{Crain2015, Bower2017, Taylor2017}, and those studies are explored by observations \citep{Martin-Navarro2018}. Taken all these studies together, the origin of the luminosity difference in SNe Ia might be related to the suggested transitions. As the average mass, metallicity, and the population age of host galaxies change with these transitions, the SN progenitor properties and environments would also change as well. This, in turn, affects the SN explosion mechanism, and therefore leads to the SN Ia luminosity difference.

\subsection{Luminosity Evolution of SNe Ia?\label{sec.conclusion.evolution}}

Since the discovery of the accelerating universe, there have been many studies concerning the luminosity evolution of SNe Ia in various ways. At earlier works, most of them used photometric information, such as the host morphology \citep{Riess1998, Schmidt1998} and the SN rise time \citep{Riess1999b}. As more and more SN Ia spectroscopic data became available, recent studies compared average spectra at high-redshift with those at low-redshift \citep{Bronder2008, Foley2008, Balland2009, Sullivan2009}. On the theoretical aspects, there are several SN explosion models to study the evolution effects \citep{Hoflich1998, Timmes2003, Kasen2009}. Interestingly, most of these studies suggested that the size of luminosity evolution effect is $\sim$0.2 mag.

\begin{figure}
\centering
\includegraphics[angle=-90,width=\columnwidth]{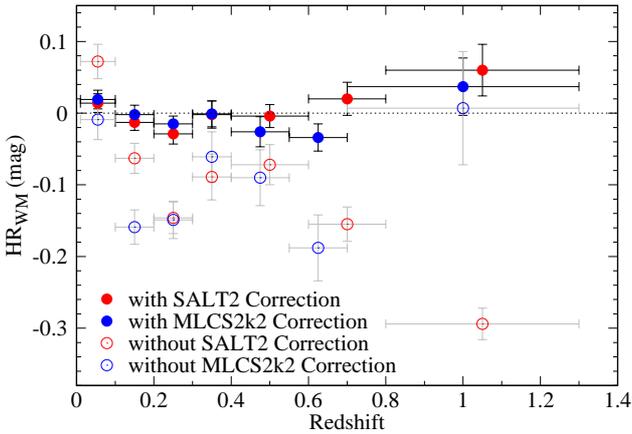}
\caption[Luminosity Evolution of SNe Ia]{Luminosity evolution of SNe Ia with respect to the redshift. Each data point is the weighted-mean of HRs in each redshift bin from the YONSEI Cosmology sample for SALT2 (red) and MLCS2k2 (blue). Filled circles are the values \textit{with} the light-curve corrections, while open circles are estimated \textit{without} the light-curve corrections. An errorbar in redshift is a bin size. A hint of the luminosity evolution is observed, but further data are required to confirm the luminosity evolution.}
\label{fig.Luminosity.Evolution}
\end{figure}

In this paper, we suggested that the stellar population age of host galaxies might be the origin of the environmental dependence of SN Ia luminosity. As the mean population age of host galaxies is known to evolve with redshift, the mean luminosity of SNe Ia also would change with redshift. In order to investigate this, we split the YONSEI Cosmology sample into several redshift bins, and calculate the weighted-mean of HRs in each bin. Figure~\ref{fig.Luminosity.Evolution} shows the evolution of the mean SNe Ia luminosity with respect to the redshift. We can see a hint of some luminosity evolution even after the standard light-curve corrections and the size of it is $<$0.1 mag, which is a smaller value than that suggested in previous theoretical studies (see above paragraph). This is because the light-curve corrections could dilute the luminosity evolution, as the model of light-curve corrections takes an average SN Ia at an average redshift \citep[see][]{Guy2007}. To examine this, we also plot the mean SN Ia luminosity \textit{without} the light-curve corrections with SALT2 and MLCS2k2 in Figure~\ref{fig.Luminosity.Evolution}. The size of the luminosity evolution \textit{without} the light-curve corrections is larger than that \textit{with} the light-curve corrections: $\sim$0.4 mag for the SALT2 sample and $\sim$0.2 mag for the MLCS2k2 sample. Interestingly, a trend of the luminosity evolution is similar until the intermediate-redshift range, while at the high-redshift range only the sample \textit{without} the SALT2 light-curve correction shows a different trend. Further data, especially at the high-redshift range, are required to confirm the luminosity evolution of SNe Ia in detail. 

In order to investigate the impact of the luminosity evolution of SNe Ia on the cosmological inference, several studies predicted the best-fit cosmological models taking into account a term for the luminosity evolution \citep[e.g.,][]{Drell2000, Linden2009, Tutusaus2017, Tutusaus2018}. The latest study of \citet{Tutusaus2018} concluded that a non-accelerated universe was able to correctly fit all the main probes if SN Ia luminosity evolution is allowed. \citet{Riess1999b} mentioned that an unexpected luminosity evolution would be sufficient to nullify the cosmological conclusions. Therefore, in order to confirm whether the luminosity evolution of SNe Ia is important or not, we require more SN data at $z>1$ (see Figure~\ref{fig.Luminosity.Evolution}), where the effect of dark energy and the luminosity evolution is distinguishable.

\subsection{Future Work\label{sec.conclusion.future}}

The current data is good enough to put the discussion for the origin and luminosity evolution of SNe Ia on the table, although more observations for SNe Ia and their host galaxies are still required.

Especially, as investigated by \citet{Kang2016} (see Section~\ref{sec.conclusion.origin}), in order to conclude the current issues on the origin of the environmental dependence and the evolution of SNe Ia at the significant confidence level, we need more, at least 100, early-type host galaxies. From ongoing SN surveys at the low-redshift range, such as the Zwicky Transient Facility \citep{Bellm2019, Graham2019}, the Korea Microlensing Telescope Network \citep{Kim2016}, and the Foundation SN Survey \citep{Foley2018}, plenty of early-type hosts will be taken for obtaining the high signal-to-noise ratio spectra to determine more reliable population ages and metallicities for host galaxies.  Furthermore, in the era of 30 m class telescopes, we would expect that we can expand the redshift range to 0.5 or above. The Dark Energy Survey \citep{Flaugher2015} SN program expects $\sim$270 early-type host galaxies to be observed up to the redshift of 1.0. At this high-redshift, a classification of galaxy morphology is another challenge. In order to more reliably select early-type host galaxies in this redshift range, we should develop or improve the method for the galaxy morphology classification, as has done before for the SDSS sample \citep[e.g.,][]{Park2005, Choi2010} and the HST Cluster SN Survey sample \citep[see][]{Meyers2012, Suzuki2012}.


\acknowledgments

We thank the referees for a number of helpful comments. This work was  supported by the National Research Foundation of Korea to the Center for Galaxy Evolution Research through the grant programs No. 2017R1A5A1070354 and 2017R1A2B3002919. Y.-L.K. acknowledges support from the European Research Council (ERC) under the European Union’s Horizon 2020 research and innovation programme (grant agreement No. 759194 - USNAC). This research has made use of the NASA/IPAC Extragalactic Database (NED), which is operated by the Jet Propulsion Laboratory, California Institute of Technology, under contract with the National Aeronautics and Space Administration. We acknowledge the usage of the HyperLeda database (http://leda.univ-lyon1.fr).




\end{document}